\documentclass[twocolumn,trackchanges]{aastex63}
\usepackage{CJK} 



\sfcode`A=1000 \sfcode`B=1000 \sfcode`C=1000 \sfcode`D=1000
\sfcode`E=1000 \sfcode`F=1000 \sfcode`G=1000 \sfcode`H=1000
\sfcode`I=1000 \sfcode`J=1000 \sfcode`K=1000 \sfcode`L=1000
\sfcode`M=1000 \sfcode`N=1000 \sfcode`O=1000 \sfcode`P=1000
\sfcode`Q=1000 \sfcode`R=1000 \sfcode`S=1000 \sfcode`T=1000
\sfcode`U=1000 \sfcode`V=1000 \sfcode`W=1000 \sfcode`X=1000
\sfcode`Y=1000 \sfcode`Z=1000

\newcommand{\js}{}
\newcommand{\etal}{{\js et~al.\/}}

\newcommand{\Msun}{\mbox{M$_\sun$}}
\newcommand{\Lsun}{\mbox{L$_\sun$}}
\newcommand{\0}{\phantom{0}}  % for spacing in tables and such

\newcommand{\sst}{{\it Spitzer Space Telescope}}
\newcommand{\s}{{\it Spitzer}}
\newcommand{\h}{{\it Herschel}}
\newcommand{\ak}{{\it Akari}}
\newcommand{\Ch}{{\it Chandra}}
\newcommand{\IRAS}{{\it IRAS}}
\newcommand{\HST}{{\it HST}}
\newcommand{\hi}{\ion{H}{1}}
\newcommand{\hii}{\ion{H}{2}}
\newcommand{\mmjy}{\hbox{$\mu$Jy}}
\newcommand{\mjy}{\hbox{mJy}}
\newcommand{\lfir}{\hbox{$L_{\rm FIR}$}}
\newcommand{\lir}{\hbox{$L_{\rm IR}$}}
\newcommand{\ergs}{\hbox{erg\,s$^{-1}$}}
\newcommand{\kms}{\hbox{km\,s$^{-1}$}}
\newcommand{\EW}{{\rm EW}}
\accepted{Mar, 2021}

\shorttitle{MIR SEDs of 16\,\micron-Selected Galaxies at $z\sim1$ }
\shortauthors{J.-S.~Huang, Y.-S.~Dai, S.~P.~Willner, et~al.}

\begin{document}
\title{A Complete 16\,\micron-Selected Galaxy Sample 
at $z\sim1$: Mid-infrared Spectral Energy Distributions} 

\author[0000-0001-6511-8745]{J.-S.~Huang}
\affiliation{Chinese Academy of Sciences South America Center for Astronomy (CASSACA),
National Astronomical Observatories, Chinese Academy
  of Sciences, Beijing 100012, China}
\author[0000-0002-7928-416X]{Y.-S.~Dai
\begin{CJK}{UTF8}{bsmi}
(戴昱)
\end{CJK}}\footnote{corresponding author: ydai@nao.cas.cn}
\affiliation{Chinese Academy of Sciences South America Center for Astronomy (CASSACA),
National Astronomical Observatories, Chinese Academy
  of Sciences, Beijing 100012, China}
\author[0000-0002-9895-5758]{S.~P.~Willner}
\affiliation{Center for Astrophysics \textbar\ Harvard \&
  Smithsonian, Cambridge, MA 02138, USA}
\author[0000-0003-4996-214X]{S.~M.~Faber}
\affiliation{University of California Observatories/Lick
  Observatory, University of California, Santa Cruz, CA 95064, USA}
\author{C.~Cheng}
\affiliation{Chinese Academy of Sciences South America Center for Astronomy (CASSACA),
National Astronomical Observatories, Chinese Academy
  of Sciences, Beijing 100012, China}
\author{H.~Xu}
\affiliation{Chinese Academy of Sciences South America Center for Astronomy (CASSACA),
National Astronomical Observatories, Chinese Academy
  of Sciences, Beijing 100012, China}
\author{H.~Yan}
\affiliation{Department of Physics and Astronomy, University of
  Missouri-Columbia, MO, USA}
\author{S.~Wu}
\affiliation{Chinese Academy of Sciences South America Center for Astronomy (CASSACA),
National Astronomical Observatories, Chinese Academy
  of Sciences, Beijing 100012, China}
\author{X.~Shao}
\affiliation{Chinese Academy of Sciences South America Center for Astronomy (CASSACA),
National Astronomical Observatories, Chinese Academy
  of Sciences, Beijing 100012, China}
\author{C.~Hao}
\affiliation{Tianjin Astrophysics Center, Tianjin Normal University,
  Tianjin 300387, China} 
\author{X.~Xia}
\affiliation{Tianjin Astrophysics Center, Tianjin Normal University,
  Tianjin 300387, China} 
\author[0000-0001-6854-7545]{D.~Rigopoulou}
\affiliation{Department of Astrophysics, Oxford University, Keble
  Road, Oxford, OX1 3RH, UK}
\author[0000-0002-4005-9619]{M.~Pereira Santaella}
\affiliation{Department of Astrophysics, Oxford University, Keble
  Road, Oxford, OX1 3RH, UK}
\affiliation{Centro de Astrobiolog\'ia (CSIC-INTA), Ctra.\ de Ajalvir,
  km~4, 28850, Torrej\'on de Ardoz, Madrid, Spain}
\author[0000-0002-4872-2294]{G.~Magdis}
\affiliation{Dark Cosmology Centre, Niels Bohr Institute, University
  of Copenhagen, Juliane Mariesvej 30, 2100, Copenhagen, Denmark}
\author[0000-0001-9197-7623]{I.~Cortzen}
\affiliation{Dark Cosmology Centre, Niels Bohr Institute, University
  of Copenhagen, Juliane Mariesvej 30, 2100, Copenhagen, Denmark}
\author[0000-0002-0670-0708]{G.~G.~Fazio}
\affiliation{Center for Astrophysics \textbar\ Harvard \&
  Smithsonian, Cambridge, MA 02138, USA}
\author{P.~Assmann}
\affiliation{Department of Astronomy, University of Concepcion,
  Concepcion, Chile}
\author{L.~Fan}
\affiliation{Department of Astrophysics, University of Science
  and Technology of China, Hefei, Anhui, China}
\author{M.~Musin}
\affiliation{Chinese Academy of Sciences South America Center for Astronomy (CASSACA),
National Astronomical Observatories, Chinese Academy
  of Sciences, Beijing 100012, China}
\author{Z.~Wang}
\affiliation{Chinese Academy of Sciences South America Center for Astronomy (CASSACA),
National Astronomical Observatories, Chinese Academy
  of Sciences, Beijing 100012, China}
\author{K.~C.~Xu}
\affiliation{Chinese Academy of Sciences South America Center for Astronomy (CASSACA),
National Astronomical Observatories, Chinese Academy
  of Sciences, Beijing 100012, China}
\author{C.~He}
\affiliation{Chinese Academy of Sciences South America Center for Astronomy (CASSACA),
National Astronomical Observatories, Chinese Academy
  of Sciences, Beijing 100012, China}
\author[0000-0003-1845-4900]{A.~Esamdin}
\affiliation{Xinjiang Astronomical Observatory, Chinese Academy
  of Sciences, Urumqi, Xinjiang, China}

\begin{abstract}
  We describe a complete, flux-density-limited sample of galaxies at
  redshift $0.8 < z < 1.3$ selected at 16\,\micron. At the selection
  wavelength near 8\,\micron\ rest, the observed emission comes both
  from dust heated by intense star formation and from active galactic
  nuclei (AGNs).  Fitting the spectral energy distributions (SEDs) of
  the sample galaxies to local-galaxy templates reveals that more
  than half the galaxies have SEDs dominated by star formation. About
  one sixth of the galaxy SEDs are dominated by an AGN, and nearly
  all the rest of the SEDs are composite.  Comparison with X-ray and
  far-infrared observations shows that combinations of luminosities
  at rest-frame 4.5 and 8\,\micron\ give good measures of both AGN
  luminosity and star-formation rate. The sample galaxies mostly
  follow the established star-forming main sequence for $z=1$
  galaxies, but of the galaxies more than 0.5\,dex above that main
  sequence, more than half have AGN-type SEDs. Similarly, the most
  luminous AGNs tend to have higher star-formation rates than the
  main sequence value. Galaxies with stellar masses
  $>$10$^{11}$\,\Msun\ are unlikely to host an AGN.  About 1\% of the
  sample galaxies show an SED with dust emission typical of neither
  star formation nor an AGN.
\end{abstract}


\section{INTRODUCTION}

Star formation rate (SFR), stellar mass, and the growth of central
supermassive black holes (SMBHs) are critical factors regulating mass
assembly in galaxies.  Star formation occurring in galaxies can be
classified in three phases \citep[e.g.,][]{daddi2010,elbaz2011}: a
main sequence in which SFR and stellar mass for most galaxies have a
redshift-dependent correlation with roughly a factor of three (but
varying with stellar mass) dispersion \citep{Davies2019}; a starburst
phase in which galaxies have SFR more than a factor of three above
the main-sequence relation \citep{Elbaz2018}; and a quiescent phase
with SFR more than a factor of three below the main-sequence
relation. 

Rapid cessation of star formation (``quenching'') is required
for massive galaxies to limit their numbers to those observed in the
local Universe \citep{faber2007,huang2003,huang2013}. Several
proposed mechanisms to quench star formation involve galaxies'
central SMBHs.  SMBH masses are linearly proportional to masses of
their hosting bulge \citep{kormendy1995, magorrian1998}, and
therefore every galaxy bulge is presumed to contain an SMBH.  The
SMBH must grow along with its host bulge to maintain the observed
linearity
\citep{rigopoulou2009,netzer2009,rosario2013,lapi2014,lanzuisi2017}. 
Indeed many observations have found a correlation between AGN accretion
luminosity and host galaxy SFR
\citep{Hao2005,Hao2008,silverman2008,madau2014,dai2018}.

A galaxy's spectral energy distribution (SED) in the rest-frame near- and
mid-infrared (NIR and MIR: 1--30\,\micron) contains rich information
about stellar mass, star formation, and AGN activity. To estimate
galaxy stellar masses, photometry at rest wavelength $\lambda\sim600$\,nm can be used,
but NIR is better \citep{bell2003,huang2013, mcgaugh2014}. Much shorter
wavelengths than 600~nm are not good because they are emitted only by
hotter stars that make up only a small fraction of the mass.  Because
stellar emission from galaxies at $z>1$ is shifted to observed NIR for
rest-frame 600~nm and to MIR for rest-frame NIR, estimating stellar mass
with only observed visible photometry becomes impossible. After the
\sst\ was launched in 2003, IRAC 3.6--8\,\micron\ photometry became the
benchmark for measuring stellar mass for high-redshift galaxies
\citep{rigopoulou2009,magdis2010,huang2013}.

Despite the complexity of galaxy SEDs, photometry at
rest-frame MIR SEDs wavelengths is often used to estimate
SFRs. Based
on a local star-forming galaxy sample, \citet{calzetti2007} argued
that rest-frame 24\,\micron\ emission arises from hot dust heated
directly by OB stars in star-formation regions, and therefore
24\,\micron\ luminosities in these galaxies are linearly correlated
with SFR. \citet{chary2001,alonso2006,calzetti2007,rieke2009} used
local star-forming galaxy samples to establish such a linear
conversion and proposed that it can apply to galaxies at high
redshift with a correct set of templates. However, estimating SFR for
galaxies at high redshift using their {\em observed} 24\,\micron\
flux densities requires an accurate K-correction. At $1<z<3$, the
MIPS 24\,\micron\ band samples rest-frame $6\,\micron
<\lambda<12$\,\micron, where there are strong spectral features such
as the PAH emission features at 6.2, 7.7, and 8.6\,\micron\ and the
silicate absorption at 10\,\micron. Some dusty galaxies at $z\sim1.4$
have such a deep silicate absorption that they have no 24\,\micron\
detection even in the deepest MIPS image but are clearly detected at
longer wavelengths \citep{magdis2011}. Therefore the MIPS
24\,\micron\ K-correction for galaxies in this redshift range can
vary substantially and is very sensitive to both redshift and the
SED. A 24\,\micron-selected sample may therefore yield a diverse
galaxy population.

The rest-frame 8\,\micron\ luminosity $L_8$ is also considered a
tracer of SFR \citep[e.g.,][]{Wu2005,Mahajan2019}. Broadband
photometry at this (rest) wavelength measures mainly PAH emission
features at 7.7 and 8.6\,\micron\ perhaps with some contribution from
the 6.2\,\micron\ feature \citep{Pahre2004}. More recently,
\citet{cortzen2019} found that the PAH emission correlates with cold
molecular gas in star-forming galaxies, but that should not be a
problem because molecular gas will be
correlated with SFR in most galaxies. Even so, PAH emission may
not trace star formation: (1) in \hii\ regions, where PAH molecules
can be destroyed by the strong UV radiation field
\citep{helou2001,houck2004, pety2005}; (2) where sources unrelated to
star formation, such as evolved stars and diffuse light, excite PAH
\citep{lidraine2002,boselli2004,peeters2004}; (3) in galaxies with
low metallicity \citep{engelbracht2005, hogg2005, galliano2005,
  rosenberg2006, shao2020}.  \citet{shao2020} found the ratio of
8\,\micron\ luminosity to SFR remains constant for galaxies with
$M_*>10^{9}$\,\Msun\ but decreases rapidly with metallicity for
galaxies with $M_*<10^{9}$\,\Msun. Despite the potential complications, \citet{elbaz2011}
found that rest-frame 8\,\micron\ luminosity $L_{8}$ has a good linear
correlation with \lir\ with $\langle \lir/L_{8} \rangle\sim5$ for
galaxies at $0<z<2$, and \citet{Mahajan2019} found $L_8$ to be a good
SFR measure for most local star-forming galaxies. Because $L_8$ is such a good
star-formation tracer, many studies have used MIPS
24\,\micron\ surveys to select star-forming galaxies at $z\sim2$
\citep{huang2007,farrah2008,desai2009,huang2009,fang2014}.

AGN emission is another uncertain factor adding to already-complicated
galaxy MIR SEDs. A bare AGN typically shows a $\nu F_\nu\approx
\rm constant$ SED in the MIR \citep{Ward1987}, much redder than
starlight. {Some PG quasars even display a silicate  emission
feature from the inner side of their dusty torus \citep{hao2005}.}
Any AGN emission shows up in the IRAC bands
\citep{lacy2004,stern2005,alonso2006} and is often the dominant MIR
component.
When the AGN component is dominant, it is
proportional to X-ray luminosity
\citep{Carleton1987,lutz2004,lanzuisi2009, stern2015}. If an AGN
component is neglected, the SFR derived from rest-frame 8~\micron\ 
luminosity or observed 24~\micron\ flux density will be overestimated.  
For galaxies with overwhelming AGN emission, MIR photometry does not measure SFR at all.

This paper presents an SED study of a complete 16\,\micron-selected
galaxy sample at $0.8<z<1.3$. It is part of series studying
rest-frame 8\,\micron-selected galaxies at $z=0.3$, 1, and 1.9 via
the observed bands of IRAC 8\,\micron, IRS peakup/\ak\ $\sim$16\,\micron,
and MIPS 24\,\micron\ \citep{huang2007,huang2009,fang2014,shao2020}.
At $z\sim1$, the PAH emission features at 6.2, 7.7, and 8.6\,\micron\
are shifted into the observed 16\,\micron\ band.  The sample will
therefore contain many star-forming galaxies but also strong
AGNs. Fitting the galaxy SEDs with a set of local templates reveals
the demography of this sample and identifies the AGNs.  A well
determined SED also permits accurate measurement of monochromatic
luminosities $L_{4.5}$ and $L_8$, thus giving an estimate of AGN
luminosity and SFR. 
The 16\,\micron\ sample is from the well-studied
extragalactic fields Extended Groth Strip (EGS), GOODS-South
(GOODS-S), and GOODS-North (GOODS-N), providing a high rate of
spectroscopic redshifts.  When spectroscopic redshifts are not
available, we use photometric redshifts, which are reliable in these
fields \citep{dahlen2013,huang2013}.
In the GOODS fields, our galaxy sample is complete to
$L_8=7.7\times10^9$\,\Lsun\ at $z=1$ corresponding to $\rm
SFR=3$\,\Msun\,yr$^{-1}$.

The structure of this paper is: \S2 describes the sample. \S3
presents MIR SEDs of star forming galaxies and AGNs in the sample. 
\S4 gives SFR estimates and
establishes the SFR--stellar-mass and SFR--AGN luminosity relations
for the sample.  \S5 is a brief summary.
Throughout the paper, we adopt the \citet{Chabrier2003} initial mass
function (IMF).  Source distances are based on a standard flat
$\Lambda$CDM cosmology with $H_0=70$\,\kms\,Mpc$^{-1}$ and $\Omega_M
= 0.30$.

\section{The 16\,\micron-Selected Sample}

\begin{figure}
%\epsscale{0.8}
\plotone{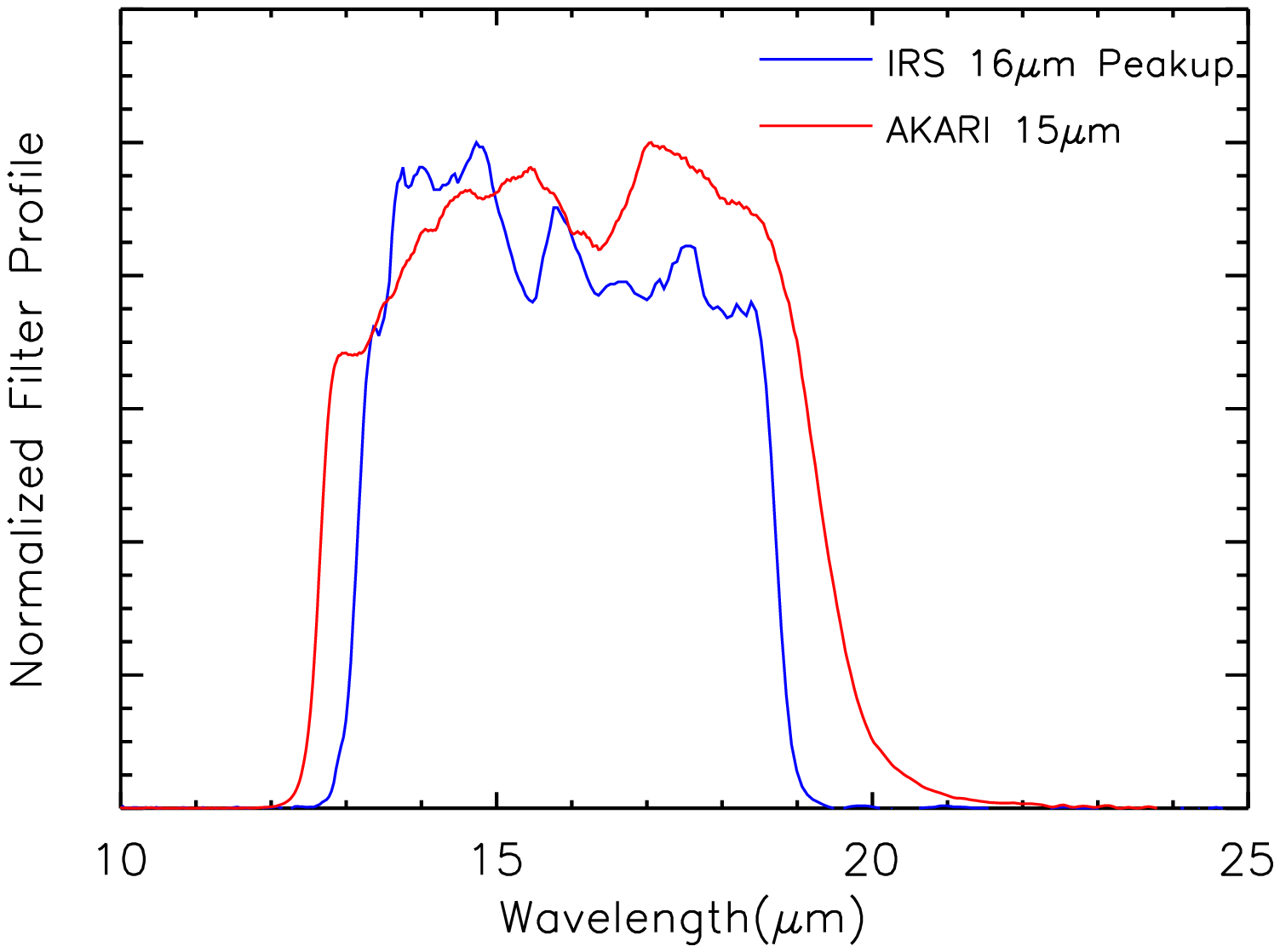}
\caption{Normalized filter transmission profiles.  Red line shows the
  \ak/IRC  15\,\micron\ profile, and blue line shows the
  \s/IRS 16\,\micron\  peakup imager profile.  The \ak\ 15\,\micron\  filter
  profile in general overlaps with but is slightly wider than the IRS
  16\,\micron\   profile.}
\label{f:fprofile}
\end{figure}

We chose a 16\,\micron-selected sample for galaxies at $z\sim1$ to
include the strong PAH emission features.  The earliest surveys were
by the {\em Infrared Space Observatory} \citep{elbaz1999,
  gruppioni2002, rodighiero2004} at 15\,\micron.  Larger and deeper
surveys came from \ak/IRC at 15\,\micron\ and the \s/IRS peakup
imager at 16\,\micron\
\citep{wada2007,burgarella2009,pearson2010,teplitz2011}.  The present
study uses photometric catalogs from the latter two
surveys. Figure~\ref{f:fprofile} shows that the \s/IRS and \ak/IRC
bandpasses have similar profiles.  \citet{teplitz2011} compared the
flux densities for GOODS-S objects detected by both instruments and
found an average 1.3 times higher flux density measured by \s/IRS at
16\,\micron\ than by \ak/IRC at 15\,\micron.  We have scaled the \ak\
photometry accordingly to its IRS 16\,\micron\ equivalent.

Our 16\,\micron\ sample\footnote{Hereafter a 16\,\micron-selected
  sample refers to the galaxy sample selected from either \ak\
  15\,\micron\ or \s/IRS-Peakup 16\,\micron\ photometric catalogs.}
has flux density limits $f_{16}>30$\,\mmjy\ in GOODS-N/S and
$f_{15}>100$\,\mmjy\ (corresponding to $f_{16}>130$\,\mmjy) in the EGS, the respective detection limits.
All three fields have extensive redshift surveys, and all
16\,\micron\ sources above our selection limits have redshifts
available, either spectroscopic or photometric.  Our final sample
comprises sources with $0.8 <z< 1.3$. This includes {\em 556 objects}
based on spectroscopic redshifts and {\em 149 objects} based on
photometric redshifts.  Table~\ref{tab:sample} summarizes the sample
characteristics.

Accurate redshifts are crucial for understanding the properties of
sample galaxies.  MIR galaxy spectra can include strong PAH emission
and silicate absorption features.  Redshift uncertainties will
therefore cause substantial uncertainties in the K-correction to
convert observed 16\,\micron\ flux densities ($f_{16}$) to rest-frame
8\,\micron\ luminositie $L_{8}$ for galaxies at $z\sim1$.  As noted
above, 79\% of the sample galaxies have spectroscopic redshifts. The
extensive multi-band photometry in our survey fields also gives
excellent photometric redshift measurements with
$\Delta z/(1+z)\sim0.03$ \citep{dahlen2013,huang2013}.  This gives
K-corrections adequate to within several percent, depending on the
exact redshift and SED of individual sources.

\begin{figure}
%\especial{0.8}
\plotone{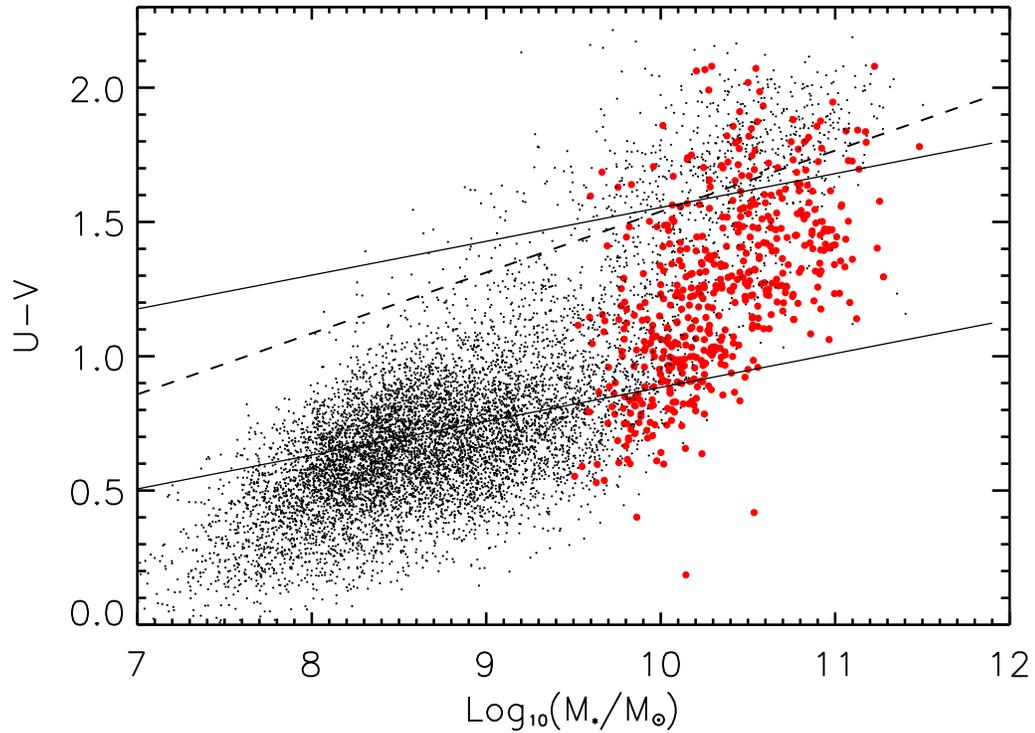}
\caption{Color--mass distribution of galaxies.  Red points indicate galaxies in
  our 16\,\micron-selected sample.  Black dots show comparison
  galaxies in the same redshift range from
  \citet{fang2018}. Horizontal coordinate represents log of stellar
  mass in solar units,
  and vertical coordinate 
  represents rest-frame $U-V$ color {corrected for dust extinction.
  The typical uncertainties in the mass estimate are $\sim$0.16\,dex \citep{mobasher15}. }
  The dashed line separates the ``red sequence''
  from the ``blue cloud'' and ``green valley''
  \citep{borch2006}. The two solid lines show upper and lower boundaries for
  the green valley for galaxies with  $\log_{10}(M_*/\Msun)>10$ at
  $z\approx1$ \citep{wang2017}.}
\label{f:ub_mass}
\end{figure}

Multi-wavelength photometry is required for this project.  We need
photometry in all four IRAC bands as well as MIPS 24\,\micron\ for
the SED fitting and classification.  The three fields have deep
\s/IRAC and MIPS coverage, and every 16\,\micron\ object was detected in all four
IRAC bands and in the MIPS 24\,\micron\ band.  There are also rich \HST\
and ground-based visible/NIR data available in these fields
\citep{huang2013,fang2018}.  We utilized the derived stellar mass
$M_*$ and SFR from CANDELS \citep{fang2018} when available.  For EGS
galaxies outside the CANDELS area, $M_*$ and SFR came from
\citet{huang2013}.  {(See~\S\ref{sec:agnsf} for more
  on the mass determination.)}
Figure~\ref{f:ub_mass} shows that
most galaxies in this sample have rest $U-V$ colors bluer than the
red sequence but at or near the red edge
of the blue cloud, i.e., they are in the ``green valley.''
The stellar masses for this sample are in the
range of $9.5<\log_{10}(M_*/\Msun)<11.5$ with a mean of $\langle
\log(M_*/\Msun)\rangle=10.3$.

All three fields have deep MIPS 70\,\micron\footnote{The FIDEL legacy
  survey is described at
  \url{https://irsa.ipac.caltech.edu/data/SPITZER/FIDEL/}.} and
\h/PACS and SPIRE data.  The GOODS fields have limiting flux
densities of 2.5\,\mjy\ at 70\,\micron, 1--3\,\mjy\ in the PACS bands,
and $\sim$10\,\mjy\ in the SPIRE bands.  Those sufficed to detect
70--80\% of the sample in at least one \h\ band
\citep{elbaz2011}. The EGS field has limiting flux densities of
2.5\,\mjy\ at 70\,\micron, 10\,\mjy\ in the PACS bands, and
14--16\,\mjy\ in the SPIRE bands. With this shallower \h\ coverage, only
40\% of the sample was detected
\citep{lutz2011,oliver2012}.  We calculated
\lir\footnote{In this paper, \lir\ means luminosity integrated
  between 8 and 1000\,\micron.} for the sample with two methods
\citep{elbaz2011}: 
{(A) convert 24\,\micron\ flux densities based on
the best-fit template {from \citet{chary2001}} to give $L_{\rm IR}^{\rm CE}$}, and (B) integrate the
SED given by \h\ and MIPS 70\,\micron\ flux densities to give
$L_{\rm IR}^{H}$.  Because every object in the sample was
24\,\micron-detected, we were able to estimate  $L_{\rm IR}^{\rm CE}$
for all objects in the sample, but only objects with FIR photometry
in at least 2 bands permit estimates of $L_{\rm IR}^{H}$.  For our
sample, $L_{\rm IR}^{\rm CE}$ are generally consistent with $L_{\rm IR}^{H}$
with scatter 0.18\,dex.

The different {sample-selection} depths in EGS and GOODS
yield slightly different demographics in luminosity classes.  In the two
GOODS fields, 72\% of objects are LIRGs, and only 2\% are ULIRGs. In the
{shallower} EGS, 85\% are LIRGs, and 6\% are ULIRGs.  Altogether our sample is
more than 70\% LIRGs, a dominant population for the star formation
rate density (SFRD) at $z\sim1$ \citep{lefloch2005}.

AGN emission can make a significant contribution to galaxy SEDs in
the MIR, and therefore it is important to measure the AGN
contribution.  In principle, almost all massive galaxies harbor an
AGN of different stages.  X-ray data can identify at least some AGNs,
and deep X-ray surveys from \Ch\ already exist in our three fields.
Depths are 800\,ks in the EGS \citep{nandra2015}, 2\,Ms in GOODS-N
\citep{xue2016}, and 7\,Ms in GOODS-S \citep{luo2017}.  These
depths detected 24 \Ch\ X-ray sources in the EGS, 59 in GOODS-N, and 46
in GOODS-S.  Only 10 sources in the EGS and GOODS-N are identified as
quasars with $L_X>10^{44}$\,\ergs.  Because of the deep \Ch\ exposure
in GOODS-S, most X-ray sources in this field have
$L_X\sim10^{42}$\,\ergs. Only three have $L_X>10^{43}$\,\ergs, and
none has $L_X>10^{44}$\,\ergs.  These are consistent with expectations
from the other fields, given the small area of GOODS-S
(Table~\ref{tab:sample}).  \citet{hickox2011} found that obscured
AGNs selected from IRAC colors \citep{lacy2004, stern2005} were equal
in number to  X-ray-selected
AGNs.  Regardless of selection method, all their AGNs had a clear
visible--MIR color segregation from non-AGN galaxies. 

\begin{figure}
%\epsscale{0.8}
\plotone{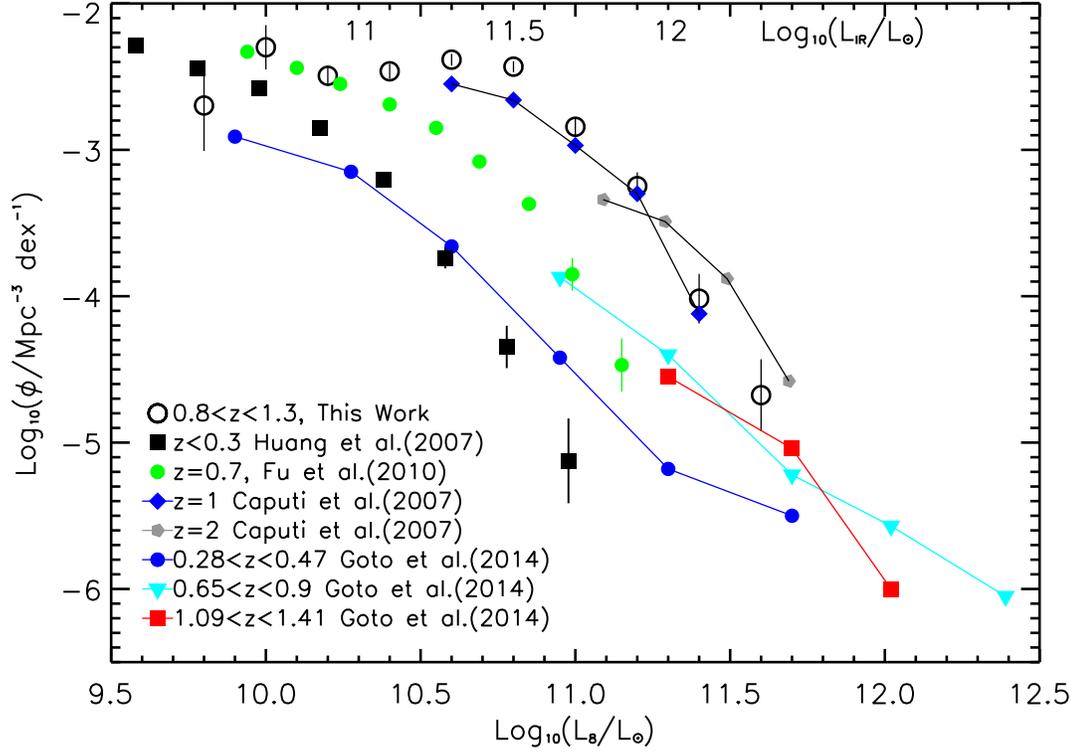}
\caption{The rest-frame 8\,\micron\ luminosity function for
  16\,\micron-selected galaxies at $0.8<z<1.3$ (open circles).  Error
  bars are based on Poisson statistics, and the corresponding
  numerical data are in Table~\ref{tab:lf8}.  Other symbols show
  rest-frame 8\,\micron\ luminosity functions from previous studies
  with redshifts indicated in the legend.  Numbers near the upper
  abscissa indicate $\log(\lir/\Lsun)$ based on the conversion
  $\log_{10}(\lir)=\log_{10}(L_8)+0.7$. Some previous studies
  \citep{matute2006, hopkins2007, huang2007, fu2010} are not plotted
  because their AGN subtraction is unclear.}
\label{f:lf8}
\end{figure}

Figure~\ref{f:lf8} shows the rest-frame 8\,\micron\ luminosity
functions for galaxies in our sample. The K-correction for $L_8$ was
derived using the best-fit SED template for each galaxy as described
in Section~\ref{s:templates}. We derived the luminosity function in
the range $10<\log_{10}(L_8/\Lsun)<11.3$ using the $1/V_{\rm max}$
method.  There are a few existing 8\,\micron\ luminosity functions at
various redshifts derived using the MIR photometry from either \ak\
or \s\ surveys \citep{caputi2007,huang2007,fu2010, goto2015}.  We
expect a strong evolution of the 8\,\micron\ luminosity function from
$z=1$ to $z=0$ because of the cosmic SFRD evolution
\citep[e.g.,][]{madau2014} and the number evolution of LIRGs
\citep[e.g.,][]{lefloch2005, goto2015}. Compared to the local
8\,\micron\ luminosity function, the luminosity function at $z\sim1$
is either 10 times brighter or 10 times higher in number density or a
mix of both.  This is consistent with the cosmic SFRD decrease since
$z\sim1$.

Our sample's total area coverage is only
694\,arcmin$^2$, and most galaxies in the sample have
$10<\log_{10}(M_*/\Msun)<11$ (Figure~\ref{f:ub_mass}).  This
combination means the derived luminosity function will be subject to
cosmic variance. This is especially the case for the 
GOODS-S field, which is roughly the
size of the Ultra-Deep Field (UDF). According to \citet{moster2011},
cosmic variances are 34\%, 25\%, and 17\% for galaxies in this mass
range in UDF, GOODS-N, and EGS respectively.  The ICRAR cosmology
calculator,\footnote{https://cosmocalc.icrar.org/} based on the work
of \citet{Driver2010} for local $M^*$ galaxies, gives 25\%, 23\%, and
18\% variance in areas equal to those of GOODS-S, GOODS-N, and EGS,
respectively. {In the three fields combined}, the cosmic variance for this sample should
be $\la$20\%.

\section{MIR SEDs of Star-Forming Galaxies and AGNs}

\subsection{Classifying the MIR  template set}
\label{s:templates}

Most galaxy MIR spectroscopic studies have been for local
galaxies. MIR spectral features from bright galaxies were first
observed using ground-based telescopes. (\citealt{moorwood1986} and
\citealt{roche1991} gave useful summaries of early work.) 
Spectra of star-forming galaxies
show strong PAH emission features at 3.3, 6.2, 7.7, 8.6, and
11.3\,\micron.  In contrast, AGNs show strong continuum emission in
this wavelength range and often show broad silicate absorption from 8
to 13\,\micron. 
With its launch in 1995, the {\it Infrared Space Observatory (ISO)}
made MIR spectroscopy possible for samples of IR-luminous galaxies
\citep{genzel1998,rigopoulou1999}.
Starting in 2003, \s/IRS spectroscopy provided even
better sensitivity than {\it ISO}, allowing spectroscopic
observations of extragalactic sources with luminosities ranging from
local dwarf galaxies to ULIRGs/HyperLIRGs (ultra- and hyper-luminous
infrared galaxies) at $z\sim2$ \citep{brandl2006,spoon2007}.  Even
now, though, spectra of distant galaxies are limited in number.

There are a few sets of broadband galaxy SED template sets 
with wavelength coverage
extending from the ultraviolet (UV) to 30\,\micron\ 
\citep[e.g.,][]{polletta2007,assef2008,assef2010,brown2014}.  
We chose the set from \citet{brown2014}, 
consisting of an atlas of 129 local-galaxy SED
templates based on spectral and photometric data observed with
grand-based telescopes, \s/IRS, and \ak/IRC.  Templates include a
wide range of SED and morphological types representative of the local
population, and they have $9 < \log(L_{\rm IR}/\Lsun)<12$ based on \IRAS\
and \h\ FIR photometry \citep{dale2012, dale2017}.

\begin{figure} 
\epsscale{0.8}
\plotone{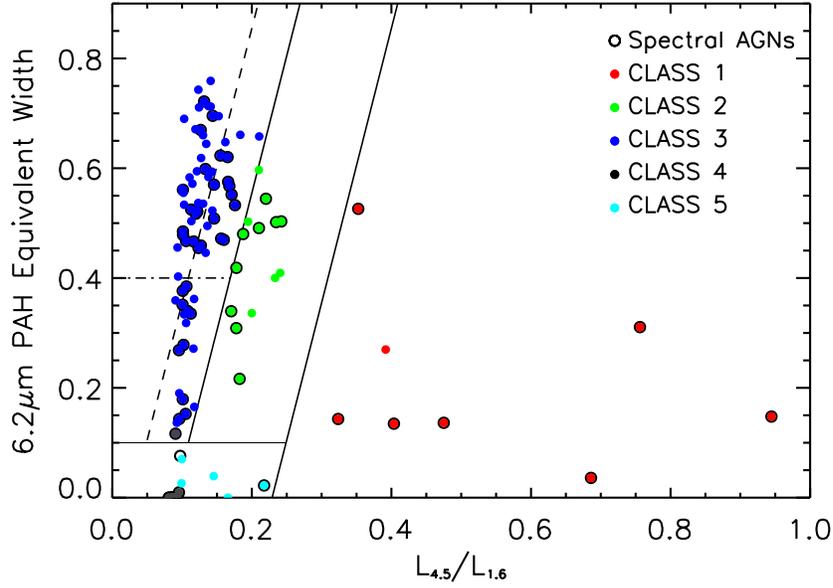}
\vspace{-14pt}
\caption{MIR classification diagram for 129 local SED templates from
  \citet{brown2014}. Points show each template's 6.2\,\micron\ PAH
  feature EW versus its $L_{4.5}/L_{1.6}$ ratio.   
  {The dashed line shows the relation of Equation~\ref{eq:ewmir},
  which was derived from Class~3 points with $\EW> 0.4$ (dash-dotted line)}.
  Points with black borders indicate templates derived from AGNs.  Lines
  show boundaries of the five template Classes. Class~1 templates are
  those with high MIR continuum and consequently low PAH EW,
  characteristic of strong AGNs.  Class~2 templates have higher EW,
  characteristic of composite AGN $+$ star-forming galaxies. Class~3
  templates are for star forming galaxies with PAH emission
  features. Class~4 are for quiescent galaxies with no dust
  emission.  They are not labeled in the figure but cluster around
  (0.09,0). Class~5 templates are for uncommon blue-compact or
  Wolf-Rayet galaxies \citep{wu2008} with relatively blue 
  {$L_{4.5}/L_{1.6}$ colors}.
  Table~\ref{tab:criteria} gives the numerical values for the
  template boundaries. }
\label{f:ew_l8l1.6}
\end{figure}

{For this work, we} divided the \citet{brown2014} templates into five distinct
Classes\footnote{We use ``Class'' instead of ``Type'' to avoid
  confusion with the well defined Type~1 and~2 AGNs.} characterized
by their predominant energy sources: star formation, starlight, and
AGNs.  In order to avoid effects of dust obscuration, Classes were
based only on rest-frame $1\,\micron\ <\lambda<30$\,\micron.  In this
wavelength range, PAH emission bands are the predominant spectral
features. The 6.2\,\micron\ PAH equivalent width (EW) is a good
 indicator of AGN strength \citep{alonso2012} with
mean 6.2\,\micron\ $\rm EW=0.24\pm0.19$ for strong AGN, $0.44\pm0.06$
for composite galaxies, and $0.52\pm0.06$ for galaxies with \hii\
spectral types. An additional classification parameter is needed
because the \citeauthor{brown2014} template set contains galaxies
with a much wider luminosity range and more diverse spectral types
than does the \citet{alonso2012} LIRG sample.  
{ We used the $L_{4.5}/L_{1.6}$ color,
as shown in Figure~\ref{f:ew_l8l1.6}, to characterize the SED shapes. 
%Our sample covers a relatively small redshift range (0.8 $ < z < $1.3), 
%thus our choice of the luminosity ratio vs luminosity ratio in the following analysis 
%will not be heavily affected by the selection effect. 
A galaxy's 4.5\,\micron\ emission may come from three components: 
stellar photospheres, star formation regions, and AGN. 
\citet{huang2007} found that when star formation activity contributes to 
the MIR continuum, it is correlated with PAH emission.  
Figure~\ref{f:ew_l8l1.6} shows typical elliptical galaxies have $L_{4.5}/L_{1.6}\approx0.085$. 
Galaxies with 6.2\,\micron\ PAH $\EW>0.4$ tend to have  higher 
$L_{4.5}/L_{1.6}$ as star formation begins to contribute. 
Galaxies with high $L_{4.5}/L_{1.6}$ and low 6.2\,\micron\ PAH EW harbor AGNs. 
A linear relation (fit with iterative $\sigma$-clipping to points
above $\EW=0.4$) gives
\begin{equation}
\EW=5\times\,(L_{4.5}/L_{1.6}-0.029)\quad.
\label{eq:ewmir}
\end{equation}}

{Figure~\ref{f:ew_l8l1.6} shows how the template Classes are
defined.  Star-forming galaxy templates have $\EW>0.1$\,\micron\
and $L_{4.5}/L_{1.6}$ within 3$\sigma$ of the corresponding value
from Equation~\ref{eq:ewmir}.  AGN templates have $L_{4.5}/L_{1.6}$
more than 10$\sigma$ greater than the Equation~\ref{eq:ewmir} value.  Templates
with offsets between 3$\sigma$ and 10$\sigma$ and $EW>0.1$ are
defined as composite.  About one-third of all templates are from
spectroscopically identified AGN hosts, but some of these show
almost no MIR spectral signature of an AGN. This happens when the
AGN luminosity is relatively small compared to the star-formation
luminosity.  Templates with no PAH emission features come from
galaxies visually classified as either elliptical or blue compact
or Wolf-Rayet type. Table~\ref{tab:criteria} specifies the
color--EW boundaries of our five Classes.}

\begin{figure} 
\epsscale{0.8}
\plotone{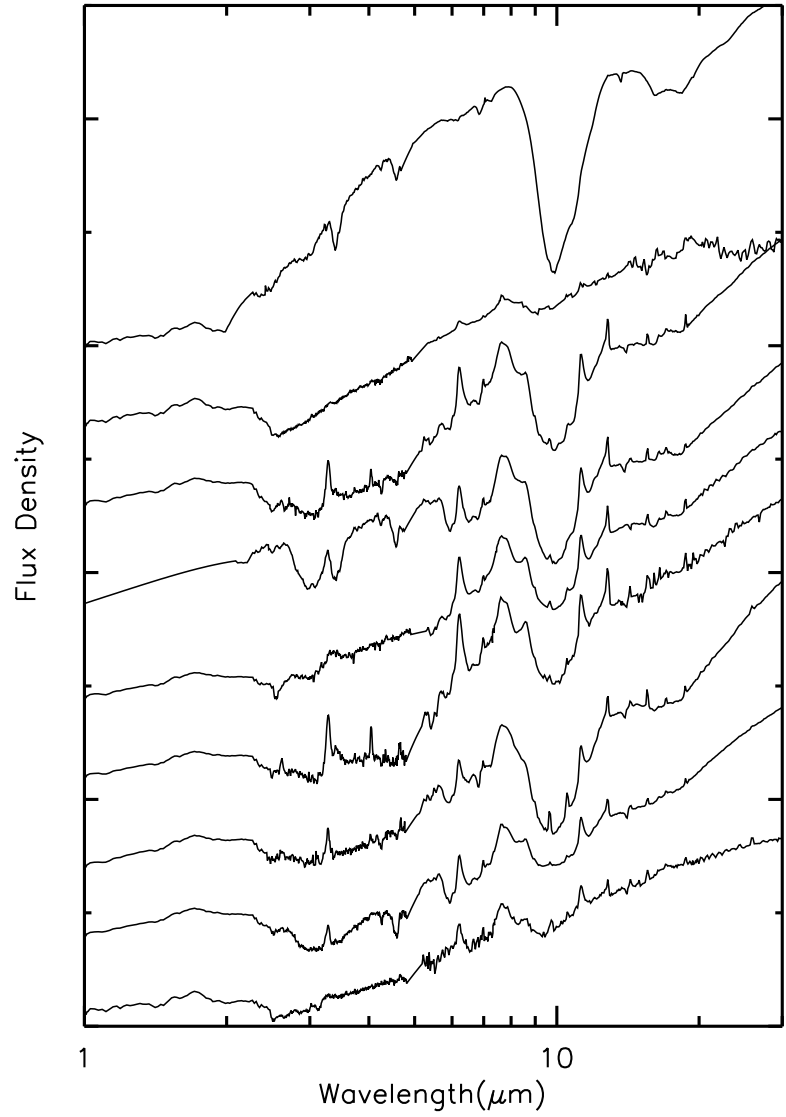}
\caption{The nine Class~1 SED templates from \citet{brown2014}.  All
  have strong continuum emission from the central AGN, and many show
  the 9.7\,\micron\ silicate feature in absorption.}
\label{f:temp_class1}
\end{figure}

\begin{figure} 
\epsscale{0.8}
\plotone{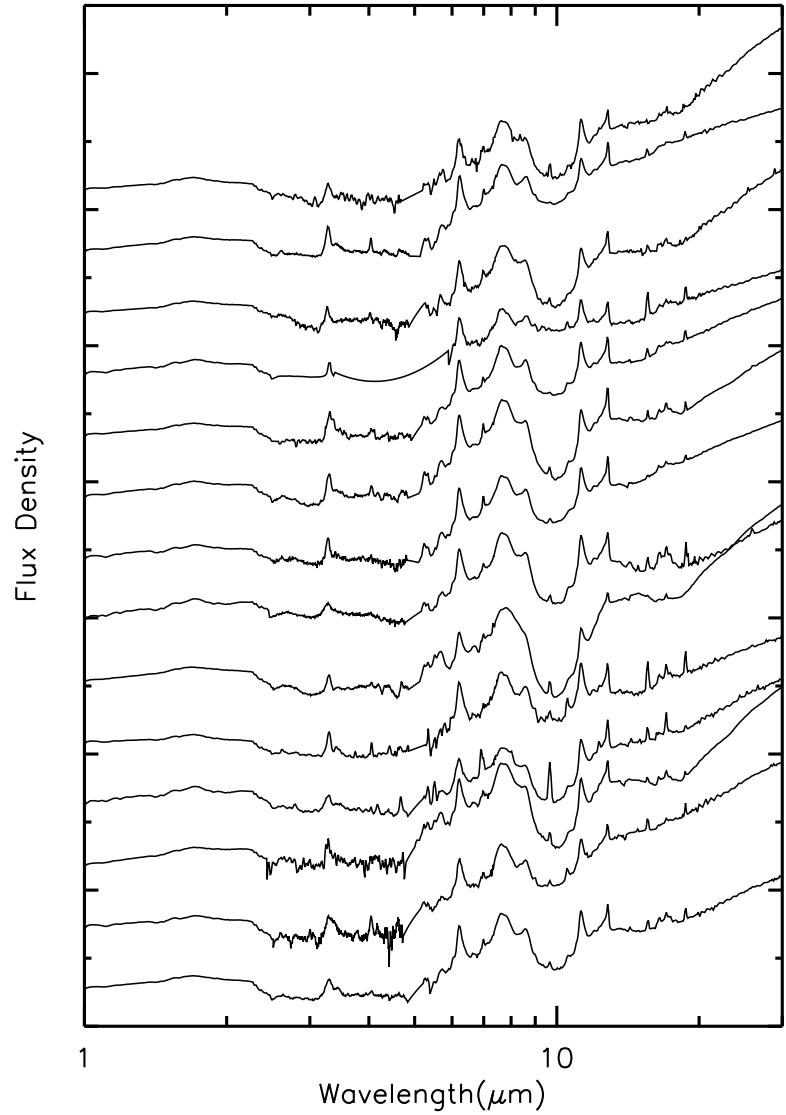}
\caption{Examples of Class~2 SED templates from \citet{brown2014}.  A
  representative 18 are 
  shown out of 20 in the complete set. The templates have composite
  SEDs with both star-forming and AGN 
  features in the MIR.}
\label{f:temp_class2}
\end{figure}

\begin{figure} 
\epsscale{0.8}
\plotone{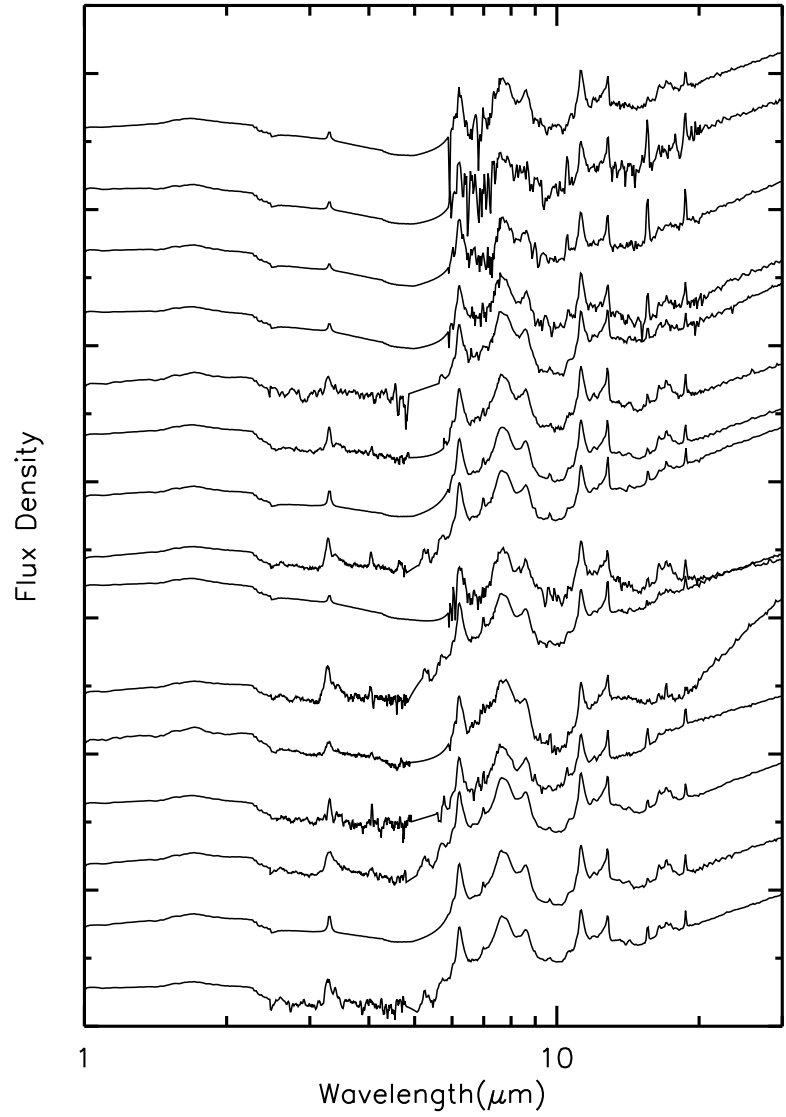}
\caption{Examples of Class~3 SED templates from \citet{brown2014}.  A
  representative 17 are 
  shown out of 68 in the complete set.  The templates
  are those of star-forming galaxies with strong PAH emission
  features.}
\label{f:temp_class3}
\end{figure}

\begin{figure} 
\epsscale{0.8}
\plotone{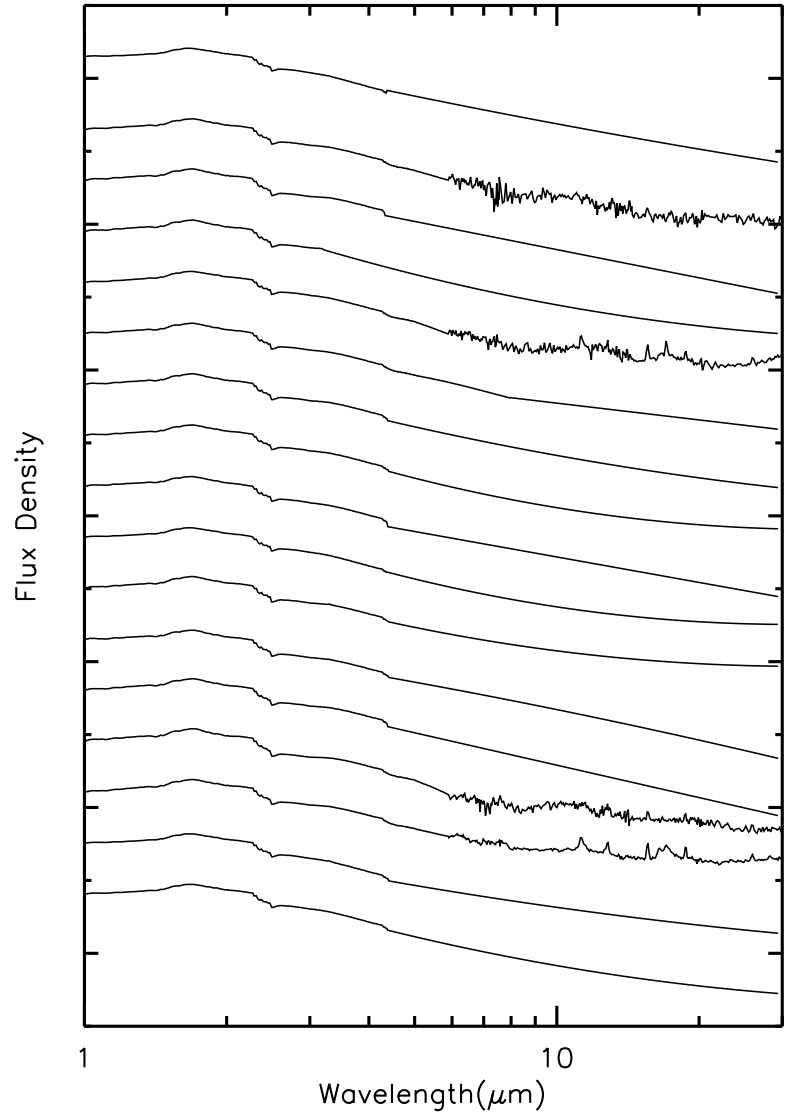}
\caption{Examples of Class~4 SED templates from \citet{brown2014}.  A
  representative 17 are 
  shown out of 23 in the complete set.  The templates
  are those of quiescent galaxies with little or no dust
  emission.}
\label{f:temp_class4}
\end{figure}

\begin{figure} 
\epsscale{0.8}
\plotone{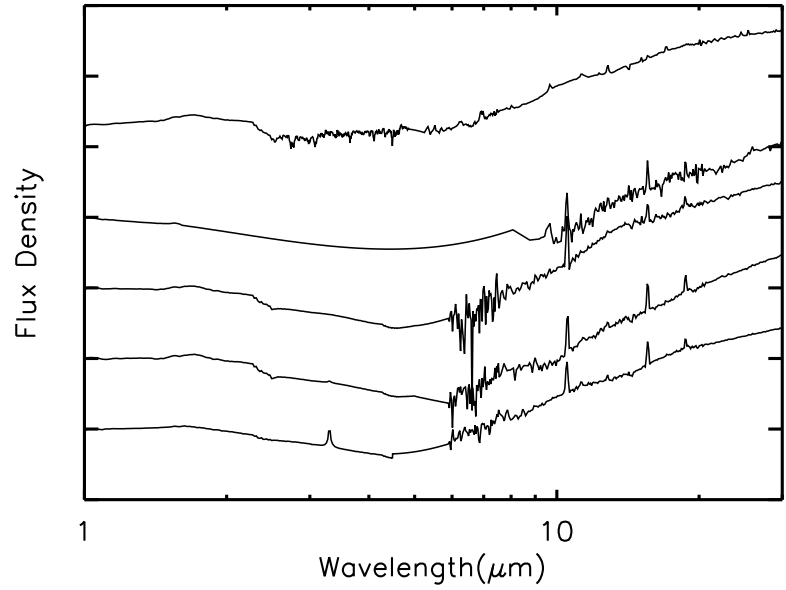}
\caption{Five Class~5 SED templates from \citet{brown2014}.
The templates are those of blue compact or Wolf-Rayet
galaxies with little or no PAH emission  
but with dust emission starting to rise steeply at $\sim$6\,\micron\
\citep{wu2008}.  
These templates have much lower
$L_{4.5}/L_{1.6}$ than the Class~1 templates.}
\label{f:temp_class5}
\end{figure}

The templates in each Class resemble each other and differ from
templates in other Classes as shown in
Figures~\ref{f:temp_class1}--\ref{f:temp_class5}. 
Class~1 templates represent AGNs and Class~2 templates represent
composite galaxies.  Star-forming templates
(Class~3) have strong PAH emission features.  Quiescent
templates  (Class~4) have nearly Rayleigh-Jeans SEDs typical of starlight.
Class~5 templates show a dust continuum
starting to rise at 5--6\,\micron\ with little PAH emission. These
nine templates (Figure~\ref{f:temp_class5}) have a strong
[\ion{S}{4}] emission line at 10.5\,\micron.  
The local galaxies in Class~5 are either young, blue, and compact or Wolf-Rayet
galaxies \citep{wu2008}, consistent with the strong [\ion{S}{4}]
line \citep{inami2013}.

When spectral information is not available, a color--color diagram
can be used to classify galaxies though with some uncertainty
as shown in Figure~\ref{f:tmp_cc}.  Galaxies with significant star
formation (Class~3) have a strong correlation between
$L_{4.5}/L_{1.6}$ and $L_8/L_{1.6}$:
\begin{equation}
L_8/L_{1.6}=11.8( L_{4.5}/L_{1.6}-0.0788)\quad,
\label{eq:nirmir}
\end{equation}
indicating that dust emission associated with star formation produces
the MIR continuum \citep{huang2007}. Strong AGN (Class~1) templates
have SEDs resembling a power law with 4.5\,\micron\ emission not much
less than that at 8\,\micron. Composite (Class~2) templates resemble
Class~3 but have slightly more 4.5~\micron\ emission than the
corresponding star-forming templates as a result of the AGN
contribution.

\subsection{SED Fitting with the Local-Galaxy Template Set}

\begin{figure}
%\epsscale{0.8}
\plotone{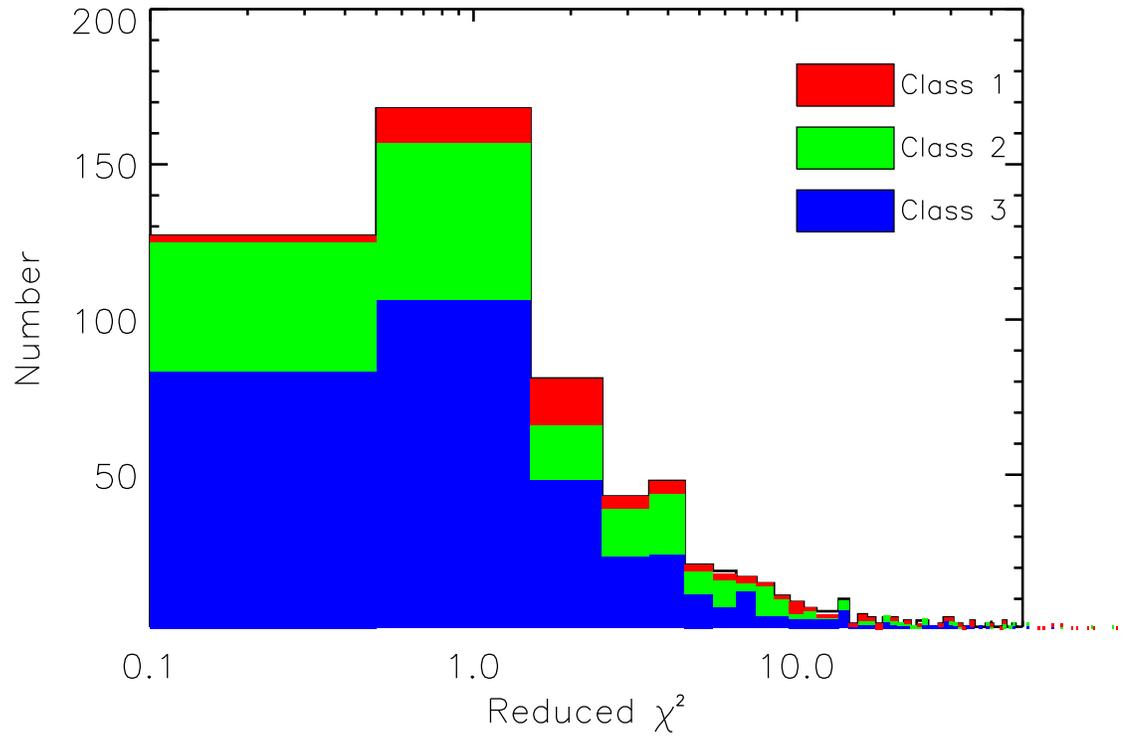}
\caption{Histogram of reduced $\chi^2$ of all SED fits for the 705
  galaxies in the sample. The color bars are counts for galaxies with
  different SED Classes. Class~5 templates are not shown but have
  $0.5<\chi^2_r<50$.}
\label{f:ki_hist}
\end{figure}

We fit SEDs only in the observed MIR wavelength range $3.6\,\micron
<\lambda<24$\,\micron.  At $z=1$, this corresponds to rest-frame
$1.8\,\micron <\lambda<12$\,\micron, where a galaxy SED has PAH
emission and silicate absorption features and is little affected by
dust extinction. 
{In order to calculate $\chi^2$ of the fits, we recalculated
  photometric uncertainties 
by putting artificial objects into the published images 
and measuring the photometric errors.
EGS has shallower IRAC and 16\micron\ depth than the GOODS fields 
and yields larger uncertainties.}    
Our SED fitting gave reduced {$\chi_r^2<2$
for 447 galaxies} in the sample, 
{and only 21 have $\chi_r^2>10$}
(Figure~\ref{f:ki_hist}). Table~\ref{tab:SED_fitting} gives the best-fit
template Class distribution for each field.
No sources in the entire sample
were fit by Class~4 (quiescent galaxy) templates. Detecting a pure-starlight
passive galaxy at 16\,\micron, even in the deep GOODS fields, would
require $M_*>3\times10^{11}$\,\Msun. Such massive galaxies are
rare, and failing to detect even one in the area surveyed is no
surprise.

\begin{figure}
%\epsscale{0.7}
\plotone{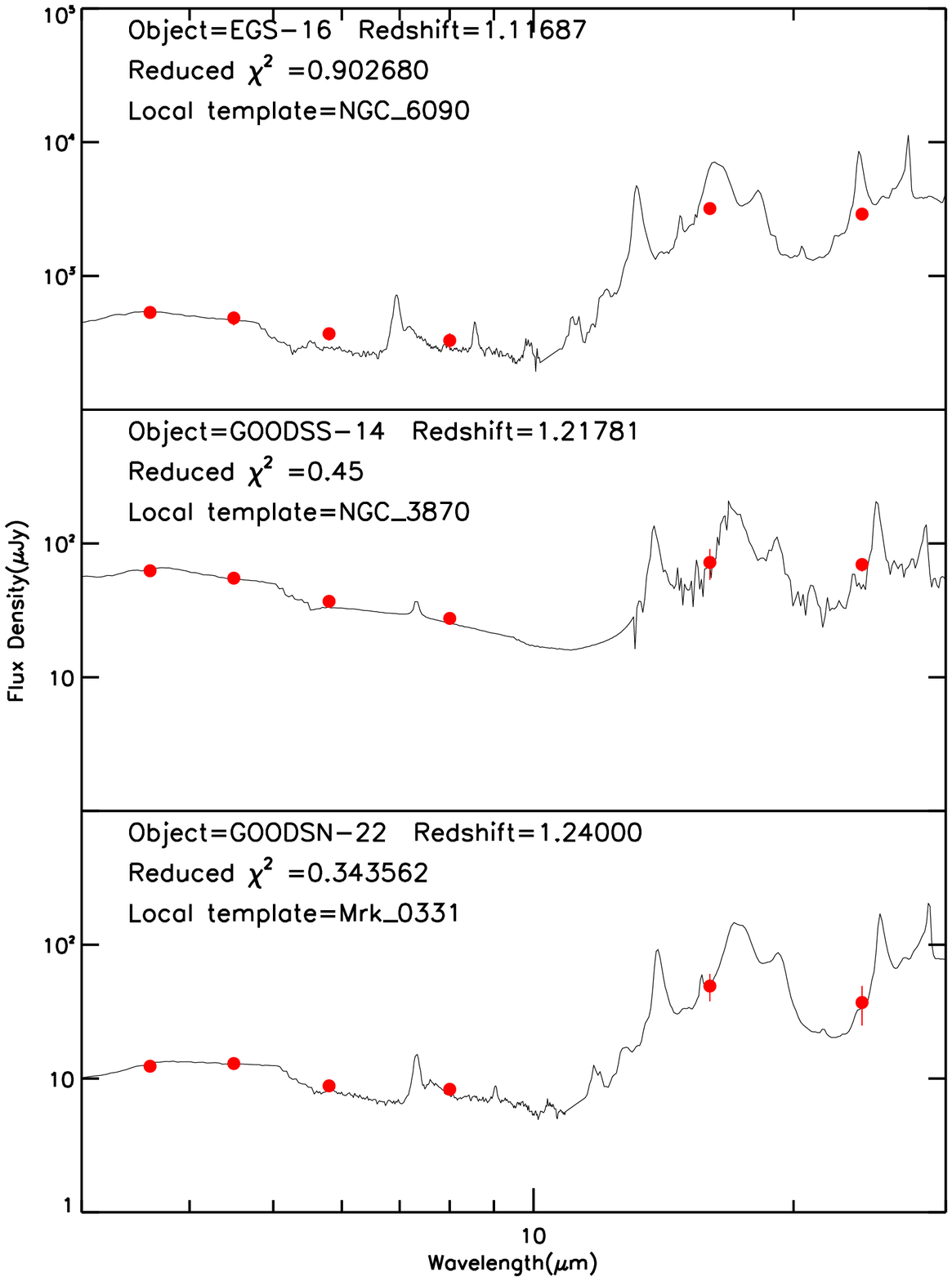}
\caption{Examples of galaxy SEDs best fit with a Class~3
  (star-forming) template.  Lines show the best-fit template, which is
  identified in each panel.  Points show the observed photometric
  data.  Wavelengths are in the observed frame, and redshifts are
  given in the panel text.}
\label{f:sf_sed}
\end{figure}

\begin{figure}
\plotone{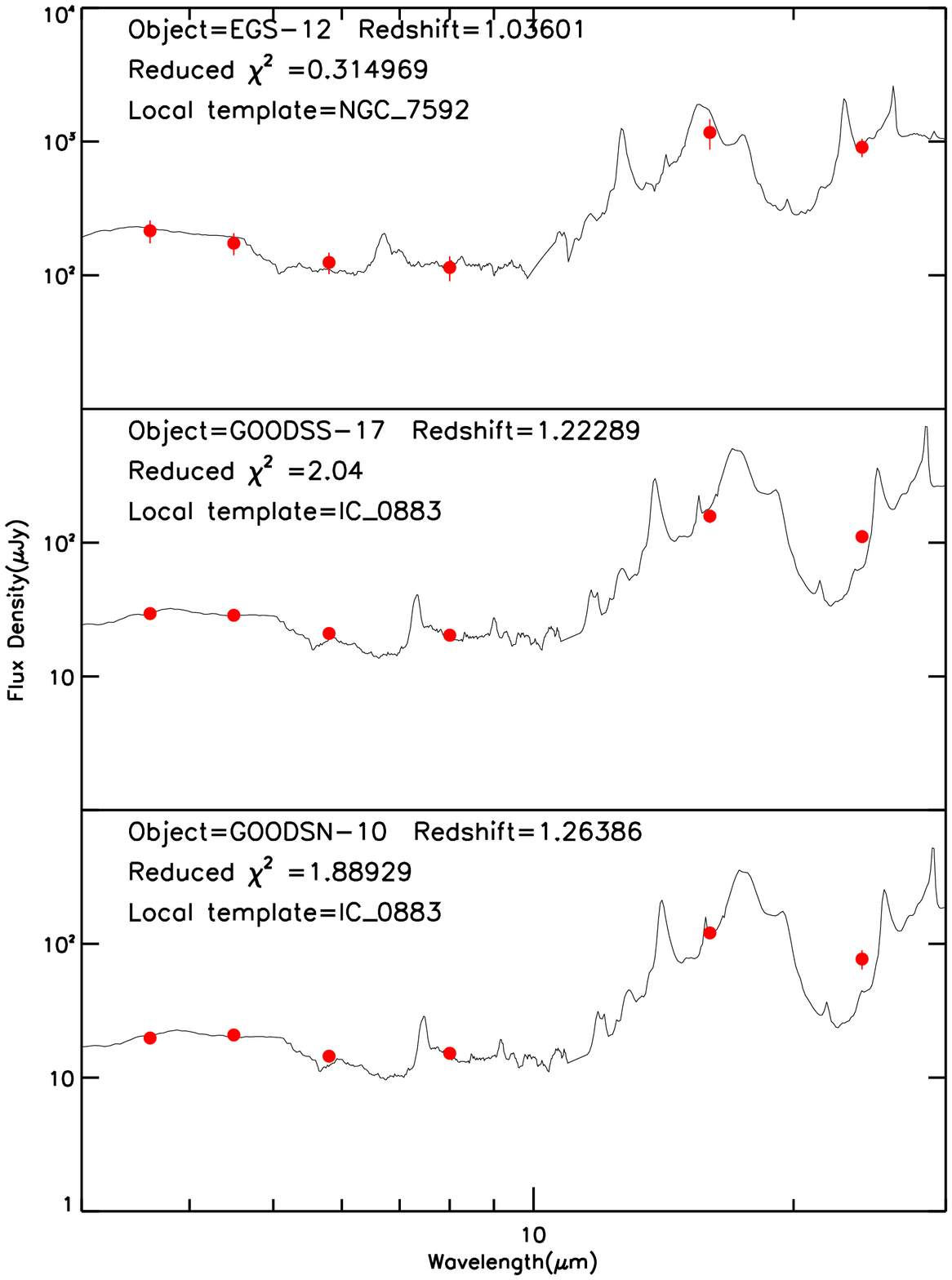}
\caption{Examples of galaxy SEDs best fit with a Class~2 (composite)
  template.  Lines show the best-fit template, which is identified in
  each panel.  Points show the observed photometric data.
  Wavelengths are in the observed frame, and redshifts are
  given in the panel text.}
\label{f:comp_sed}
\end{figure}

The best fits are for star-forming galaxies, as shown by the lower
$\chi_r^2$ values in Figure~\ref{f:ki_hist}.  The Class~3
template set covers a wide range of colors and PAH EWs as shown in
Figures~\ref{f:ew_l8l1.6} and~\ref{f:tmp_cc}, and therefore it is
not surprising that a good template can nearly always be
found. Figures~\ref{f:sf_sed} and~\ref{f:comp_sed} show examples of
good fits with Class~3 templates. 

\begin{figure}
\plotone{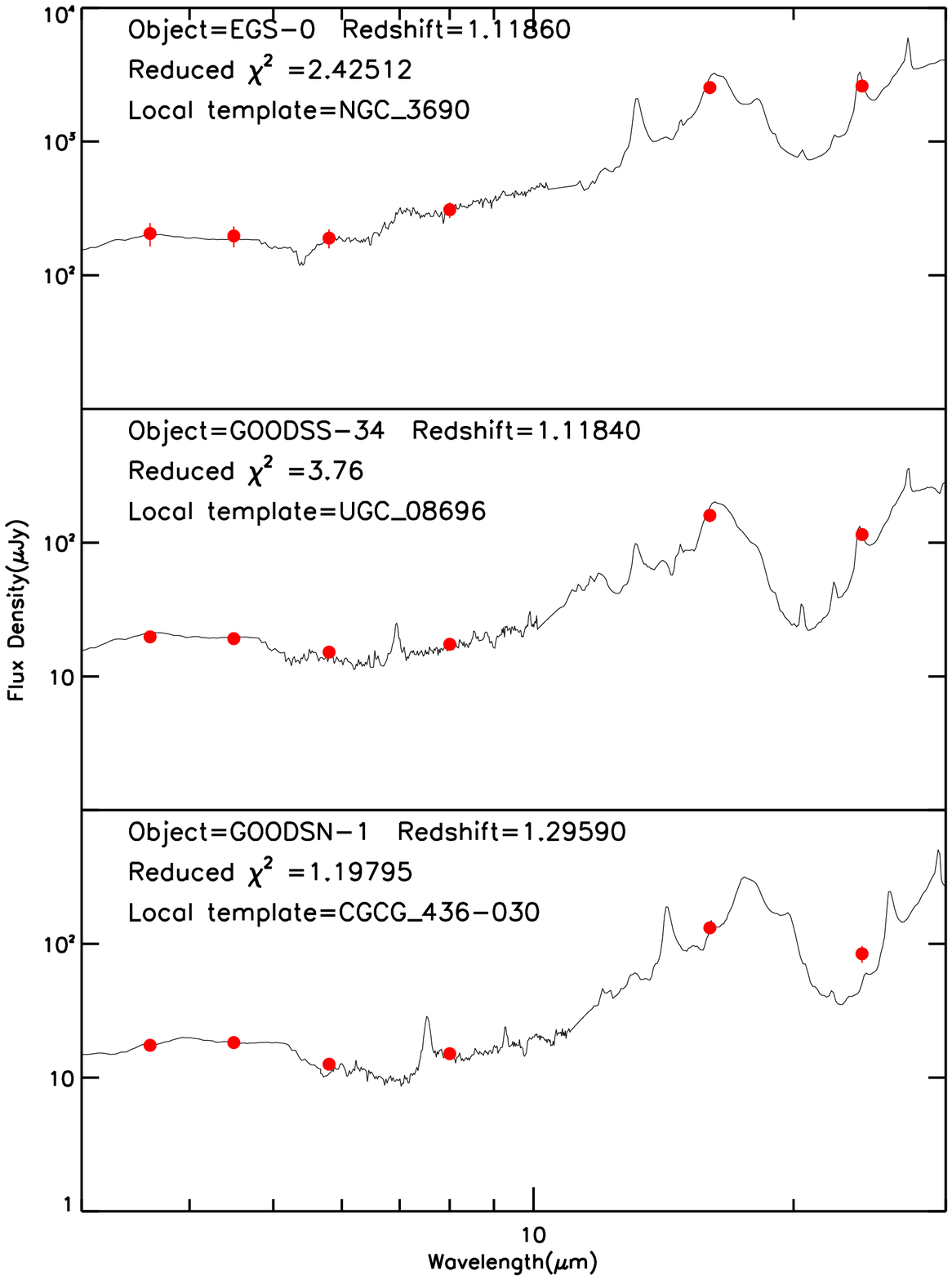}
\caption{Examples of galaxy SEDs best fit with a Class~1 (AGN)
  template.  Lines show the best-fit template, which is identified in
  each panel.  Points show the observed photometric data. Wavelengths
  are in the observed frame, and redshifts are
  given in the panel text.}
\label{f:agn_sed}
\end{figure}

\begin{figure}
\epsscale{0.7}
\plotone{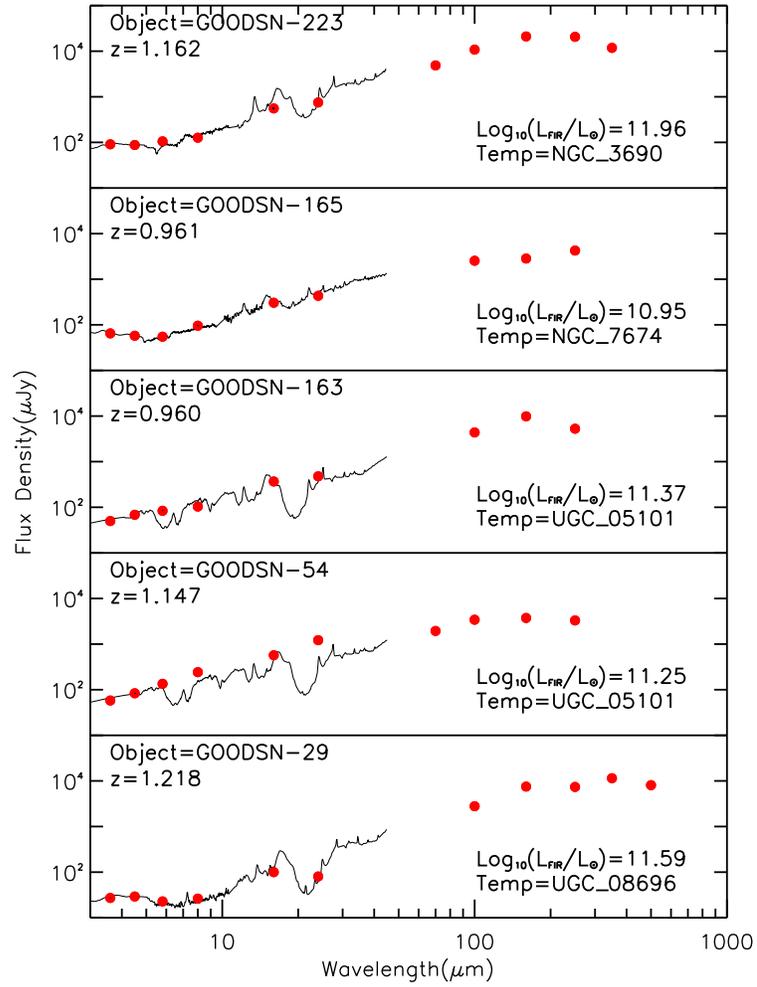}
\caption{Full IR SEDs for galaxies with dominant AGN contribution in
  the MIR (Class~1). 
  Only galaxies with positive detections in at least three \h\ bands
  are shown. Lines show the templates, and points show the observed
  data. Photometric uncertainties are smaller than the point
  sizes.  Wavelengths are in the observed frame, and redshifts are
  given in the panel text.}
\label{f:full_agn_sed}
\end{figure}

Fitting SEDs for galaxies with Class~1 (AGN) templates yields
slightly higher reduced $\chi^2$ than for star-forming
galaxies. Examples are shown in Figure~\ref{f:agn_sed}.  
{While AGN variability could be a factor at the 0.1\,mag level
  \citep{Kozlowski2016},}
two related
factors likely contribute more to the higher $\chi^2$.  One is that our
$z\approx1$ AGNs are much more luminous than local ones and
may not be represented in the local templates. The other is that
there are only nine Class~1 templates in the set, and these may not
represent the full range of AGN SEDs even among local galaxies.  A further
contributing factor is that the AGN galaxies generally have high
signal-to-noise ratios, and therefore even minor deviations from the
templates will give large $\chi^2$. 
Figure~\ref{f:tmp_cc} shows the
wide range of color space the AGNs map out and the paucity of
templates within that space. Figure~\ref {f:temp_class1} shows that
the templates include not only a range of continuum slopes but also
multiple absorption and emission features, e.g., water absorption
at 3.1\,\micron, PAH emission at 3.3\,\micron, and the bare
carbon absorption feature at 3.4\,\micron\ \citep{imanishi2001}.
Nine templates are simply not enough to cover the full range of
parameters. An example is shown in Figure~\ref{f:full_agn_sed}, where
the local UGC~5101 template includes lots of features but still deviates
from the observed GOODSN-54 SED.

{ 
With only six photometry points for each object used in template
fitting,  
multiple templates might fit an observed SED 
within the uncertainties. 
To evaluate such degeneracies, we
  compared the best and the second-best templates for each
  galaxy.
We defined a $\chi_r^2$ limit as $\exp(-\chi_1^2)/\exp(-\chi_2^2)<1.5$, 
where $\chi_i$ are the reduced $\chi$ for the best and second-best
templates, respectively.
There are 262 galaxies within this limit.
In 170 of them,
both templates belong to the same Class.  
For 86 objects, the two templates belong to adjacent Classes, either
Class 1 and Class~2 or Class~2 and Class~3. 
Only six objects have one of the two best-fit templates  from Class~1
and the other from Class~3.  
All of these six are in the EGS field, where 
larger photometric errors may contribute.   
In summary, 
most cases of degenerate template-fitting results 
are due to the similarities between the templates 
as shown in Figures~\ref{f:ew_l8l1.6} and~\ref{f:tmp_cc}, and 
the majority of our sample have a clear SED identifications.}

\begin{figure}
\plotone{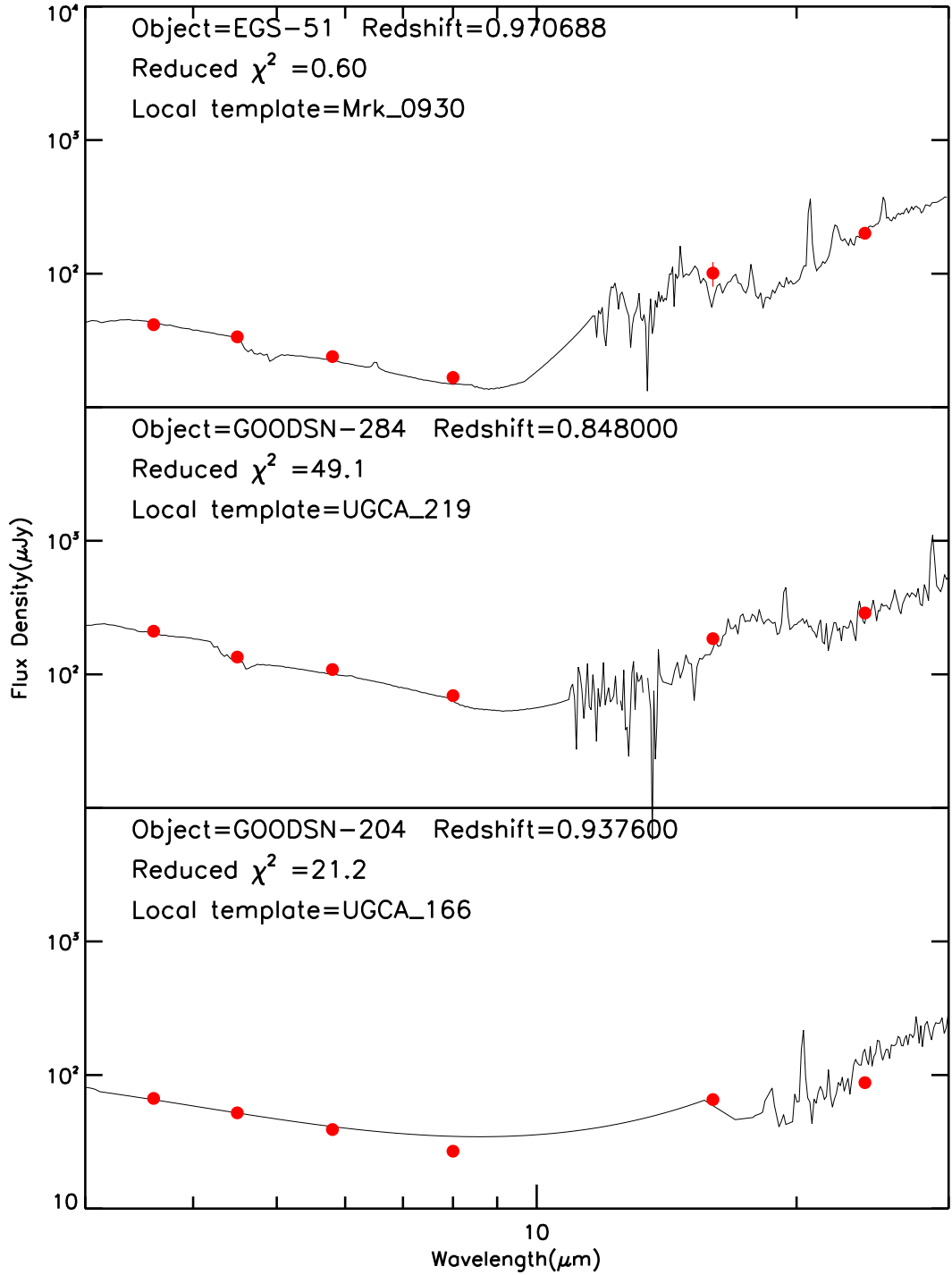}
\caption{Examples of galaxy SED best fit with Class~5 (Wolf-Rayet
  or blue compact) templates.  Lines show the best-fit template,
  which is identified in each panel.  Points show the observed
  photometric data. Wavelengths are in the observed frame, and redshifts are
  given in the panel text. Our sample
  has only 10 objects of this type.}
\label{f:wr_sed}
\end{figure}

The fitting procedure identified nine objects in the sample with
weak or absent 8\,\micron\ PAH in their SEDs but red continua near that
wavelength.  The best fits to these
objects are SEDs of UGCA~166, UGCA~219, UGCA~410, and Mrk~930. These
Class~5 templates show no or extremely weak PAH features \citep{wu2008},
but unlike normal AGNs whose strong continuum emission starts around
3\,\micron, their hot dust emission starts around rest-frame
6\,\micron\ (Figure~\ref{f:wr_sed}).  Local galaxies with this MIR SED
are typically Wolf-Rayet or blue compact galaxies with
$M_*<10^9$\,\Msun.  The lack of PAH emission is a result of either
low metallicity \citep{engelbracht2005, shao2020} or vigorous star
formation with intense UV radiation destroying PAH molecules.  The
existence of OB stars and intense UV radiation in local galaxies of
this type is supported by the detection of the [\ion{S}{4}]
$\lambda$10.54\,\micron\ and \ion{He}{2} $\lambda$4686\,\AA\ lines in
their spectra. Ionization potentials are $\sim$35\,eV and $\sim$25\,eV
respectively, demanding the presence of hard UV photons to produce
the lines.  Recent \h\ and ALMA observations of blue compact galaxies
UGCA~166 and SBS~0335$-$052 found FIR SED peaks between 20 and
60\,\micron\ \citep{hunt2014}, indicating high dust temperatures in
these systems.  This is consistent with a strong UV radiation from
compact star-formation regions.  Previous studies concluded that
these two galaxies are very young with $M_*\sim10^6$\,\Msun\
\citep{houck2004,wu2008,hunt2014}. However, the nine objects in our
sample fitting Class~5 templates have $M_*>10^{10.5}$\,\Msun. They
are therefore unlikely to have low metallicity or be in the early
stages of star formation. The galaxies show no strong X-ray emission
or excess continuum emission at rest-frame 4.5\,\micron, making AGN
an unlikely cause for the absence of PAH. Further study of this
population is needed.

\subsection{AGNs in the 16\,\micron-Selected Sample}
\label{sec:excess}
Our SED results identify that about 15\% galaxies in the sample are
best fit by local AGN templates, and another 32\% have  composite
SEDs with both AGN and star-formation contributions. Confirmation of
AGNs through spectroscopy is challenging, and we therefore resorted
to X-ray observations \citep{nandra2015,xue2016,luo2017} for AGN
confirmations.  
%We adopted $L_{\rm 2-10\,keV}$ as $L_X$ and converted $L_{\rm 0.5-7\,keV}$ in 
%\citet{xue2016,luo2017} to $L_X$ with $L_X=0.721L_{\rm 0.5-7\,keV}$. 
About 18\% of the whole 16\,\micron\ sample are X-ray
detected. Table~\ref{tab:X-sample} gives the X-ray depth of each
field and the percentages of X-ray detections for each field in each Class. 
In all fields, Class~1 sources were detected at a much higher
rate than other Classes, consistent with the AGN classification.
However, even with the 7\,Ms depth in GOODS-S, fewer than half the
presumed AGNs were X-ray detected. 

\begin{figure}
\plotone{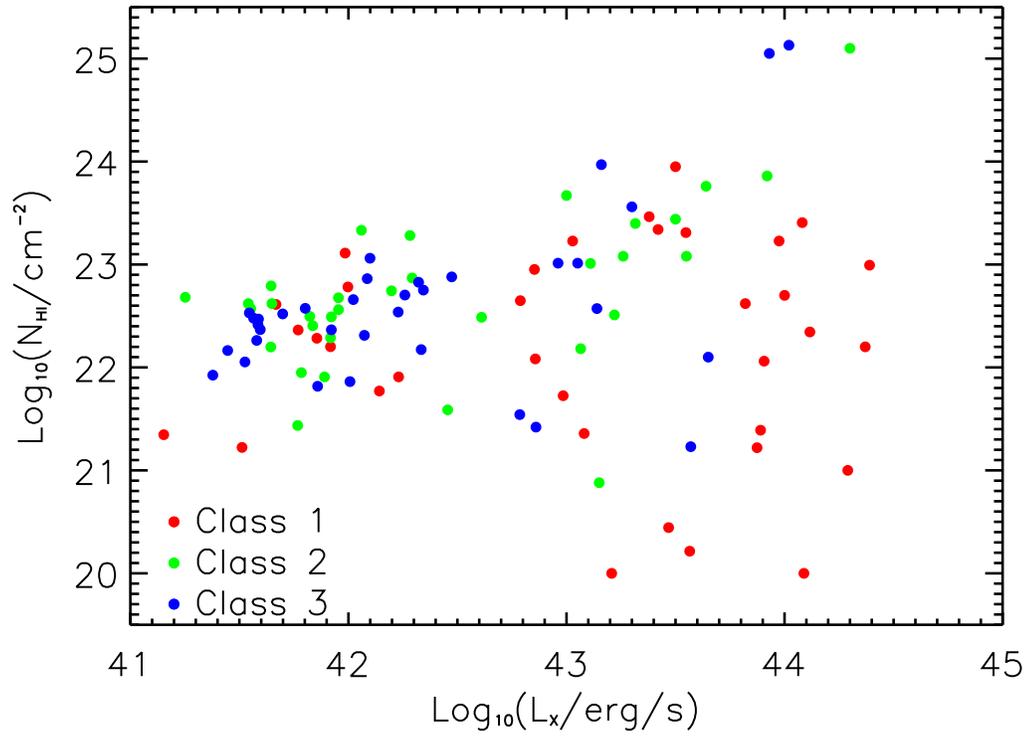}
\caption{\hi\ column densities for all X-ray sources in the sample. 
Colors indicate SED Class as indicated in the legend.
%Most have a column density $N_{\rm H}>22$ cm$^{-2}$, suggesting  that
%they are dusty AGNs.
%Most sources with high X-ray luminosity also have Class~1 SED. 
The three X-ray sources with $L_X\sim10^{44}$\,\ergs\ and not Class~1 are 
at the top of the plot. }
\label{f:lx_nh}
\end{figure}

\begin{figure}
\plotone{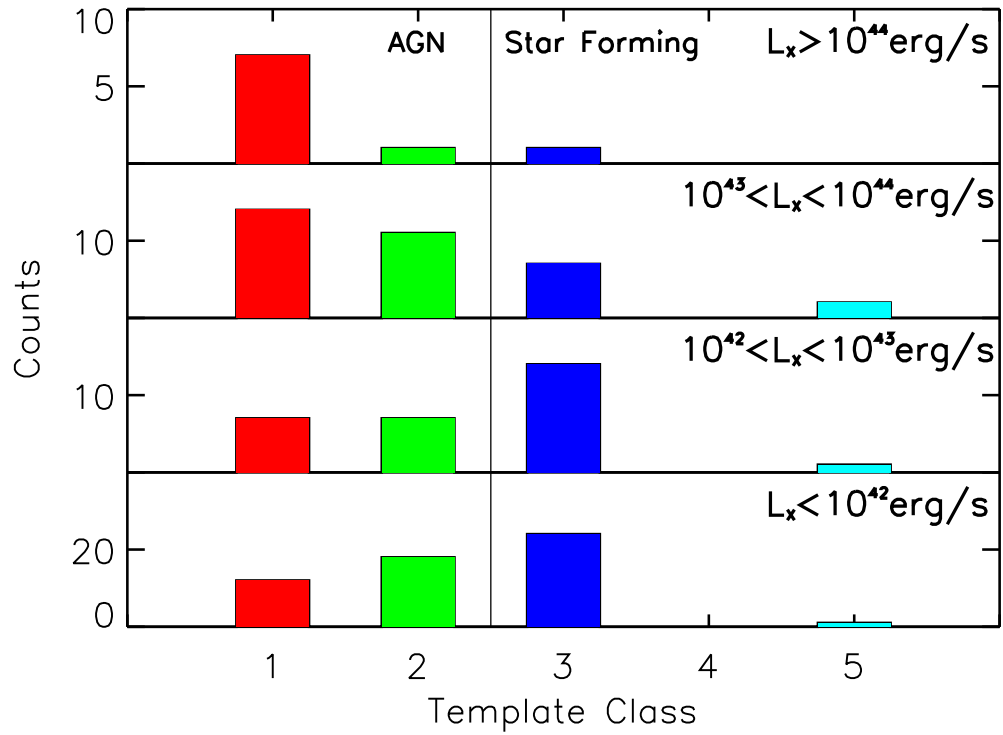}
\caption{Histograms showing SED Classes for \Ch\ X-ray sources in our
  16\,\micron-selected sample. Rows show galaxies in different bins
  of X-ray luminosity. The thin vertical line separates Classes~1
  and~2, which show MIR evidence of an AGN, from other Classes, which
  show only star-formation features in the MIR. {X-ray luminosities
  are from 2 to 10\,keV; see note to Table~\ref{tab:X-sample}.} }
\label{f:xray_temp_hist}
\end{figure}

\citet{nandra2015,xue2016,luo2017} derived physical properties
  from the X-ray SEDs including X-ray luminosities and obscuration
  expressed as gas column density $N_{\rm H}$. 
Figure~\ref{f:lx_nh} shows that about 70\% of the X-ray AGNs in this sample 
have $N_{\rm H}> 10^{22}$\,cm$^{-2}$, 
which corresponds to visual extinction $A_V\approx5$\,mag
\citep{Valencic2015}, i.e., a dusty AGN. 
Three X-ray sources have
$N_{\rm H} \ga 10^{25}$\,cm$^{-2}$, qualifying them as Compton-thick. 
They have apparent $L_X\ga10^{44}$\,\ergs, in the classical AGN range, 
but none of these three objects has a Class~1 SED, and two are
Class~3. With such large extinction, the true X-ray luminosity is
uncertain and may be lower than estimated.  A more important factor
may be the extinction, which corresponds to $A_V\approx500$\,mag and
$A_{4.5\,\micron}\approx23$\,mag
\citep{Hensley2020}. Figure~\ref{f:lx_nh} shows other, less extreme
examples of X-ray-luminous objects with Class~3 SEDs.  While the
extinction amounts are uncertain, AGNs can be so well hidden that MIR
SEDs reveal no trace of them.
Figure~\ref{f:xray_temp_hist}
shows the distribution of SED Classes for four ranges of X-ray
luminosity. Almost all X-ray sources with $L_X >10^{44}$\,\ergs,
i.e., the X-ray QSOs, have Class~1 SEDs. In contrast, most X-ray
sources with $L_X <10^{42}$\,ergs\ have a Class~3 SED, indicating
that they are predominantly star-forming galaxies.  X-ray sources
with intermediate luminosity show a range of Classes including many
composites.  Overall the percentage of objects with Class~1 SEDs
increases with X-ray luminosity, showing a good correlation between
the X-ray luminosity and MIR SED Class.

\begin{figure}

\plotone{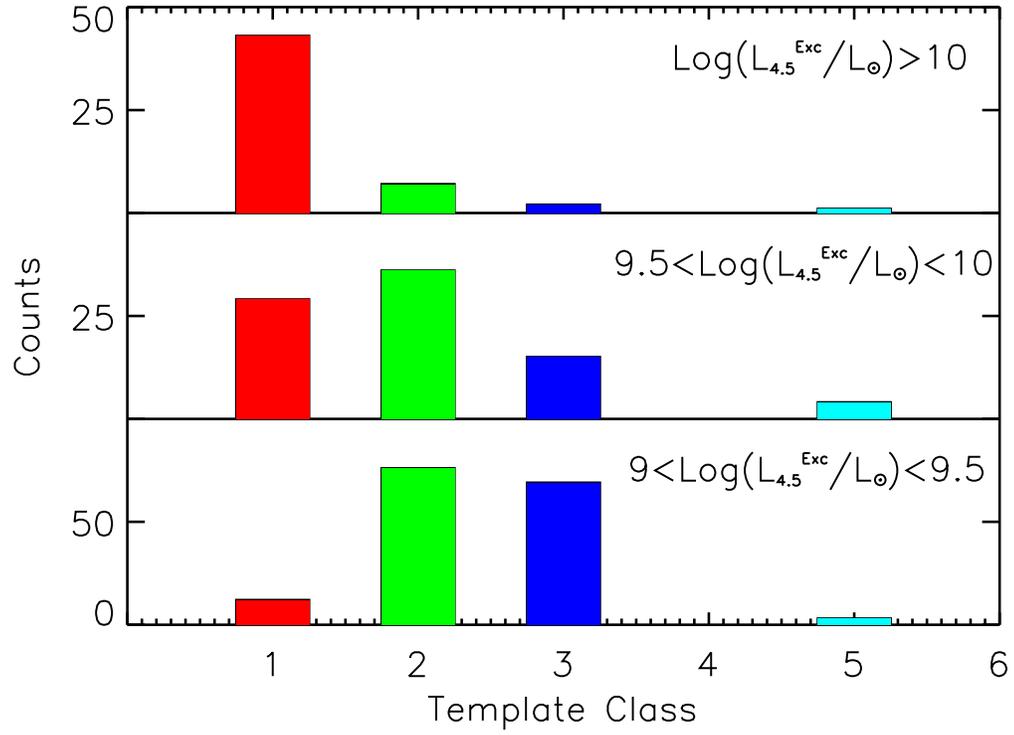}
\caption{Histograms showing SED Classes for galaxies with excess
  4.5\,\micron\ luminosity as defined by
  Equation~\ref{eq:lexcess}. Rows show galaxies in different bins of
  $L_{4.5}^{\rm Exc}$. }
\label{f:l4.5_temp_hist}
\end{figure}

\begin{figure}
\plotone{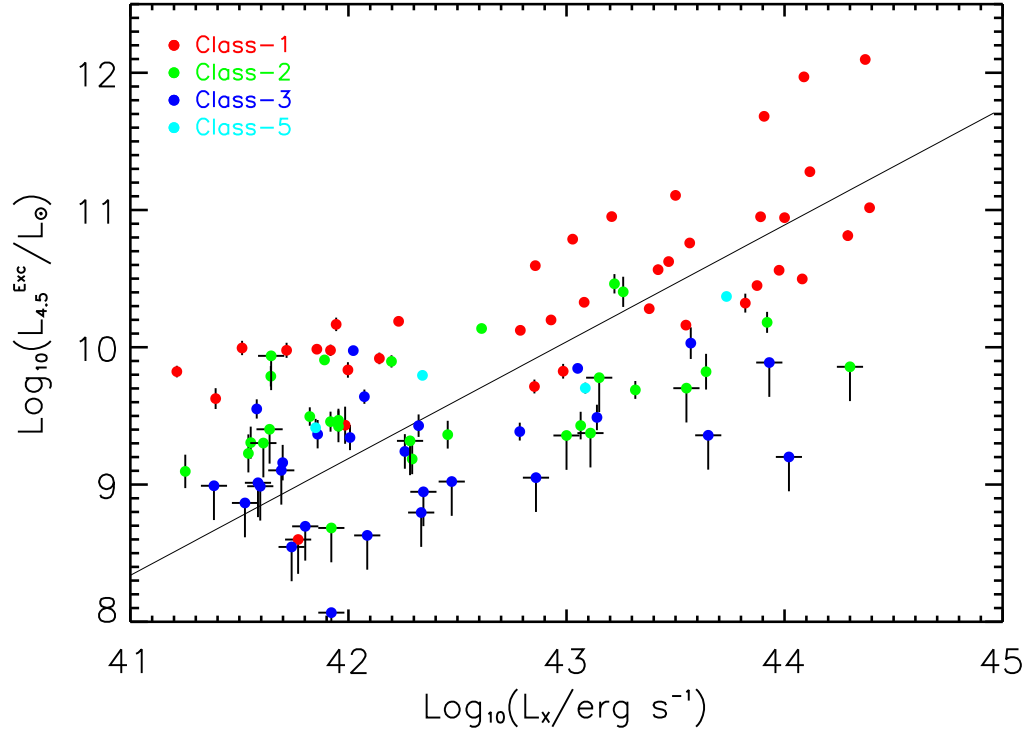}
\caption{Excess 4.5\,\micron\ luminosity (Eq.~\ref{eq:lexcess}) as a
  function of X-ray luminosity.  Points are color coded by template
  Class as shown in the legend. The line shows the best linear fit
  to the data (Eq.~\ref{eq:lxcor}).}
\label{f:l4.5_lx}
\end{figure}

Our SED fitting also yields accurate measurement of MIR luminosities
characterizing both star formation and AGN. Several studies showed a
correlation between AGNs' MIR and X-ray luminosities
\citep[e.g.,][]{Carleton1987,lutz2004, lanzuisi2009, fiore2009,
  stern2015, suh2019}.  This correlation was found for AGNs with
$L_X>10^{42}$\,\ergs, which have strong MIR continuum emission
\citep{stern2015,dai2018}, and is the same for Type~1 (broad-line)
and Type~2 (narrow line) AGNs \citep{suh2019}.  Luminosity at
rest-frame 6\,\micron, $L_{6}$ was often used to represent the MIR
luminosity in these previous studies. Most of the X-ray sources in
our sample also have PAH emission in the MIR bands.  This is
consistent with many Class~1 templates (Figure~\ref{f:temp_class1})
and all the Class~2s and~3s
(Figures~\ref{f:temp_class2},~\ref{f:temp_class3}) and is probably a
result of the 16\,\micron\ sample selection favoring sources having
7.7\,\micron\ emission features.  PAH emission being present means
that $L_6$ could be contaminated by the 6.2\,\micron\ feature. We
therefore used rest-frame 4.5\,\micron\ luminosity as a MIR measure of AGN
accretion power.  $L_8$ and $L_{4.5}$ were directly derived from each
object's SED.  These luminosities come from both star formation and
AGN. As explained in the Appendix, the excess luminosity at
4.5\,\micron, attributed to an AGN, is
\begin{equation}
L_{4.5}^{\rm Exc}=
  L_{4.5}-L_{4.5}^{\rm SFR}=\epsilon(11.8L_{4.5}-L_8+0.93L_{1.6})\quad,
\label{eq:lexcess}
\end{equation}
where $\epsilon\approx0.093$. Objects with $L_{4.5}^{\rm
  Exc}/\sigma_{4.5}^{\rm Exc}>3$ are those with a significant
excess. (Flux density uncertainties at both 8 and 16\,\micron\
contribute to $\sigma_{4.5}^{\rm Exc}$, the uncertainty of $L_{4.5}^{\rm Exc}$.)  It is
not surprising that most $L_{4.5}$-excess objects have an SED
best fit by Class~1 templates, as shown in
Figure~\ref{f:l4.5_temp_hist}.  
{There is a good correlation between
$L_{4.5}^{\rm Exc}$ and $L_X$ for X-ray sources in our sample, as
shown in Figure~\ref{f:l4.5_lx}.}  The best fit is
\begin{equation}
\log_{10}(L_{4.5}^{\rm Exc}/\Lsun)=0.85\log_{10}(L_X/10^{42}\,\ergs)+9.19\quad.
\label{eq:lxcor}
\end{equation}
X-ray sources at $L_X<10^{42}$\,\ergs\ could be powered purely by
star formation, but a few of these have high $L_{4.5}^{\rm Exc}$ and
Class~1 SEDs, indicating their AGN nature. These may be
X-ray-obscured.  The correlation between $L_X$ and $L_{4.5}^{\rm Exc}$ suggests that the excess MIR luminosity is a
signature of the active nucleus and can be used as a measure of AGN
luminosity.

\begin{figure}
\plotone{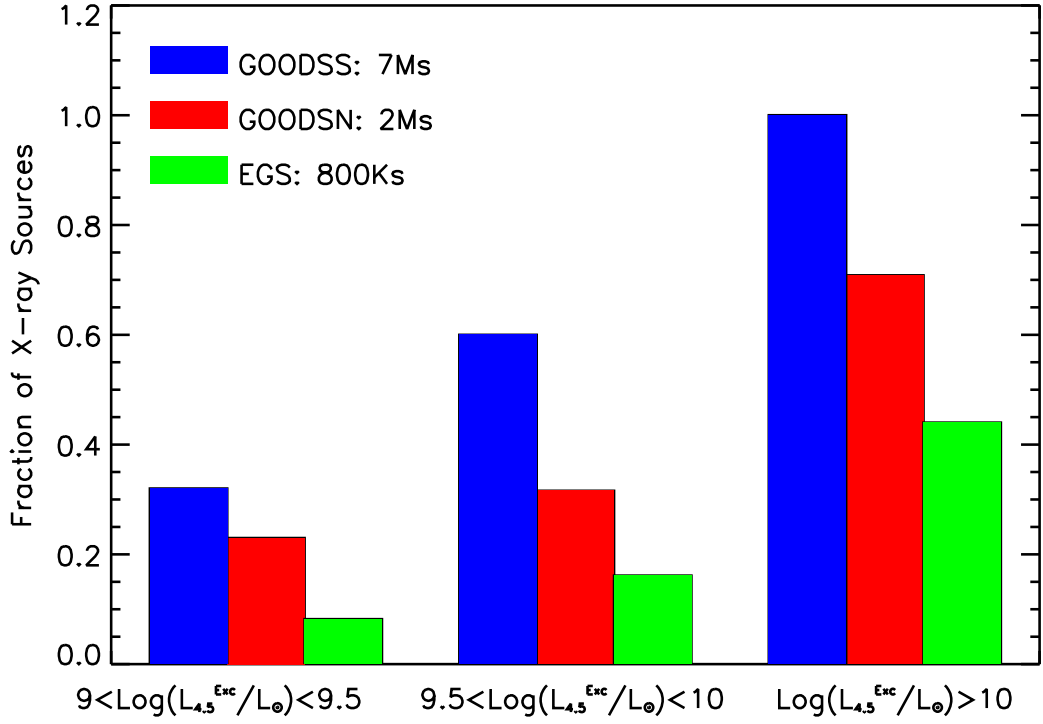}
\caption{Histograms of X-ray-detected fraction of sources with
  4.5\,\micron\ luminosity excess.  Colors show different fields,
  which have differing depths of X-ray observation as indicated in
  the legend.  Three sets show different bins of $L_{4.5}^{\rm Exc}$
  (Eq.~\ref{eq:lexcess}). }
\label{f:xray_percent_excess}
\end{figure}

There are many galaxies in the sample with strong $L_{4.5}^{\rm Exc}$ but
no X-ray detection.  The 3$\sigma$ and 5$\sigma$ limits are roughly
at $\log_{10}(L_{4.5}^{\rm Exc}/\Lsun)=9.5$ and 10.  According to
Equation~\ref{eq:lxcor}, the two limits roughly correspond to
$L_X=10^{42.1}$ and $10^{43}$\,\ergs, respectively. AGN at these
X-ray luminosities should have been detected by \Ch, but only half of
them were.  This is consistent with at least half of all AGN being
obscured in X-rays \citep{Gilli2007,Hickox2018,Lambrides2020}.  We did
not find any significant difference 
between the MIR-excess-selected objects with and without X-ray detection,
but as shown in Figure~\ref{f:xray_percent_excess}, the percentage of
X-ray-detected sources increases with exposure time and with
luminosity $L_{4.5}^{\rm Exc}$.  This trend suggests that the
MIR-excess targets without X-ray detections are X-ray-obscured, at
least in the GOODS fields.  (In the EGS, the relatively shallow X-ray
depth may also play a role.)
% was also found in comparison of MIR and optically selected
%AGNs \citep[e.g.,][]{dai2014}.
Selecting on the basis of MIR excess $L_{4.5}^{\rm
  Exc}/\sigma_{4.5}^{\rm Exc}>3$ gives 140 objects. Of these, 63
have a Class~1 SED, 45 have a Class~2 SED, and 32 have a Class~3 SED.
In the first category, all the Class~1 {\em templates} have a
  MIR excess, but with the available $S/N$, the simple $L_{4.5}^{\rm
    Exc}$ calculation finds only about 60\% of the individual objects
  fitting the AGN templates.
The 32 Class~3 objects with detectable MIR excess are 9\% of all Class~3
objects.  Even though 
  their MIR emission is powered mainly by star formation, these
  objects show a detectable amount of AGN emission.  There might be
  another 4\% or so that would show an excess if higher $S/N$
  observations were available.
The overall number of AGNs found by MIR excess is
 about double the number of $L_X>10^{42}$\,\ergs\ AGNs in
  this sample (Table~\ref{tab:X-sample}).  The combination 
of MIR and X-ray selection yields a more complete AGN sample than
X-ray selection alone, as has
been found before \citep{hickox2011}.

\section{Star Formation in the 16\,\micron-Selected Galaxies}
\subsection{SFR Estimators from the UV to FIR}

\begin{figure}
\epsscale{1.0}
\plotone{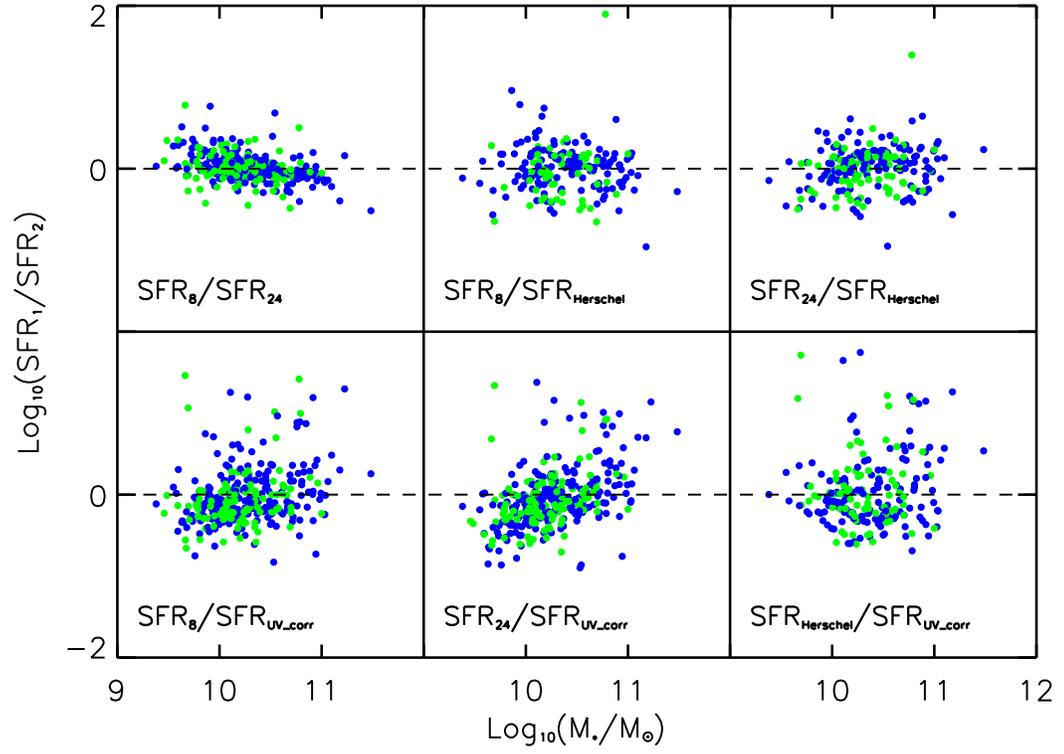}
\caption{Comparison of SFR estimators as a function of stellar
  mass. Text in each panel identifies the ratio plotted.  Only galaxies with
  Class~2 (green points) and Class~3 (blue points) SEDs, which should be relatively unaffected by
  AGNs, are shown.  Galaxies with significant $L_{4.5}^{\rm Exc}$ are
  omitted for the same reason.  Horizontal dashed lines show
  equality.}
\label{f:sfr_sfr}
\end{figure}

One goal of this project is to compare various SFR estimators for
galaxies.  Our sample has luminosities available in the UV, MIR, and
FIR, all of which can be used to estimate SFR.  For this work,
SFR$_{\rm UV}$ was calculated from UV--visible SED modeling of CANDELS data
\citep{grogin2011},
including dust correction \citep{fang2018}. SFR$_{8\,\micron}$ was
based on $L_8$ (rest frame as
derived from the 16\,\micron\ photometry) converted to SFR according
to the \citet{elbaz2011} conversion. SFR$_{24}$ was calculated using the {\em observed}
24\,\micron\ flux density and an appropriate template
\citep{chary2001}.  SFR$_{\rm FIR}$ was based on $L_{\rm IR}$ from \h\
FIR SEDs.   We excluded galaxies with
significant $L_{4.5}^{\rm Exc}$ to avoid confusion by AGN
contamination.  \citet{fang2018} found SFR$_{\rm UV}$ and SFR$_{24}$ to be
consistent for $z<1.5$ galaxies.
Figure~\ref{f:sfr_sfr} shows that all SFR estimates
yield consistent results on average, but SFR$_{\rm UV,corr}$ shows
considerably more scatter with the other three indicators than any of
them shows with another.
%Among the three \lir\ estimators, SFR based on $L_{IR}^{Herschel}$
%has a relative large scatter due to lower sensitivity in the
%\h\ FIR bands.  SFRs based on $L_{8}$ and $L_{IR}^{24}$ have
%the best correlation and suggest more accurate SFR estimates.  At the
There is a slight trend for SFR$_{24}$ to be lower than the other
SFRs at the lowest masses.  Some galaxies have very low SFR$_{\rm
  UV}$, probably because the modeling underestimated dust
extinction. Though there are arguments that PAH emission may vary
when tracing SFRs due to metallicity difference \citep{shao2020} or
heating from evolved stars \citep{crocker2013}, the correlation
between SFR and PAH luminosity has held up albeit with scatter
\citep{treyer2010} or some systematics at low and high luminosity
\citep{Mahajan2019}.  \citet{elbaz2011} found SFR$_{8\,\micron}$ to
be better than SFR$_{24}$ at $z > 2$.  Our results show that $L_8$ is
a good SFR estimator for this LIRG sample at $z\sim1$, indicating
that the PAH emission is associated with star-forming regions in
these LIRGs.

\subsection{Star Formation in AGNs}

\begin{figure}
\plotone{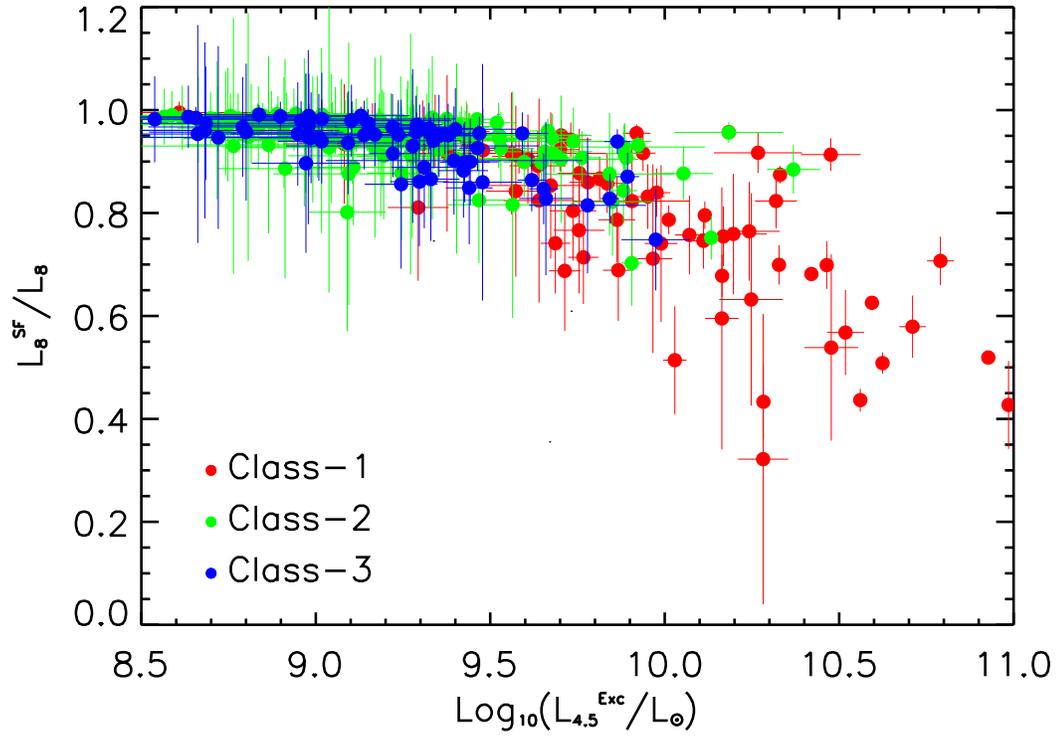}
\caption{Fraction of star formation luminosity in $L_8$ as a function
  of AGN luminosity as measured by $L_{4.5}^{\rm Exc}$. Points are
  color coded by SED type. Most galaxies with $L_{4.5}^{\rm
    Exc}>10^{9.5}$ have median $L_8^{\rm SF}/L_8=0.85$, implying that
  $L_8$ is usually a good measure of SFR. }
\label{f:lagn_sfrfra}
\end{figure}

As shown above, AGNs tend to produce MIR emission, which will cause
overestimated SFR if not excluded.  The appendix shows how to solve
a set of linear equations to separate the components into $L_8^{\rm
  SFR}$ and $L_{4.5}^{\rm Exc}$.  For most galaxies, the AGN
correction from $L_8$ to $L_8^{\rm SFR}$ and the star-formation
correction from $L_{4.5}$ to $L_{4.5}^{\rm Exc}$ are modest.
Figure~\ref{f:lagn_sfrfra} shows that for most
objects, an AGN contributes at most 10--20\% of $L_8$.  However, for
$L_{4.5}^{\rm Exc}>10^{9.5}$\,\Lsun, $L_8$ will overestimate SFR for
a significant
fraction of galaxies, nearly all of them having Class~1 SEDs. In the
whole sample, 20 galaxies have $L_8^{\rm SFR}/L_8<0.5$, and a few
have $L_8^{\rm SFR}$ consistent with zero.  For these objects,
neither $L_8$ nor $L_{24}$ is a good SFR estimator.  The FIR
luminosity, \lfir, being less affected by the AGN emission
\citep{dai2018}, should however still be a valid SFR measure.  All but six
of the 20 galaxies have enough FIR detections to give
SFR. Figure~\ref{f:full_agn_sed} shows SEDs for five examples with at
least three \h\ detections.  Their \lir\ qualifies them as LIRGs. The
overall picture confirms the correlation between $L_8$ and SFR for
all galaxies \citep{elbaz2011}, including AGNs at $z\sim1$. Our study
refines the SFR estimates and allows studying the relation between
star formation and AGN (\S\ref{sec:agnsf}).

\subsection{Co-Evolution of Galaxies and Central SMBHs}
\label{sec:agnsf}

Galaxy mass assembly is correlated with central SMBH growth.  This is
supported by the observed correlation between the galaxy-bulge
stellar mass and central SMBH mass \citep[e.g.,][]{magorrian1998,
  kormendyho2013}.  This co-evolution scenario implies correlation
between SFR and the SMBH's accretion rate as calculated from $L_{\rm
  AGN}$ \citep[e.g.,][]{netzer2009, rosario2013, dai2018, yang2019}.
Several studies found that AGNs, especially with
$L_X>10^{44}$\,\ergs, show a good correlation between SFR and $L_{\rm
  AGN}$ \citep[e.g.,][]{lutz2010, rosario2012, chen2013,chen2015,
  hickox2014, azadi2015, xu2015,dai2018,yang2019}. 

{Practical estimates of the stellar mass for AGNs can be complicated.
A strong AGN has significant emission in both the visible and NIR
bands, and this emission must
be subtracted when estimating  host-galaxy stellar mass. 
Stellar masses for galaxies in the CANDELS fields 
were derived using SEDs \citep{mobasher15}. 
This method in principle fits and subtracts the AGN contribution but
with some uncertainty.
Most X-ray AGNs in this sample have large \hi\ column densities, 
suggesting they  are dust-obscured in visible light. 
According to \citet{silva2004,aird2018}, an AGN with
$N_{H}>10^{22}$\,cm$^{-2}$ has  a
negligible contribution to its host SED in the visible bands.
MIR-selected AGNs without X-ray detection are also likely to be dusty. 
The sample has only two AGNs with $L_X>10^{44}$\,ergs\ and $N_{\rm
  H}<10^{22}$\,cm$^{-2}$.  
Yet their stellar masses are already on the lower end, at $\log M_*=
10.14$ and 10.45\,\Msun, respectively. It therefore seems unlikely that
AGNs are causing large overestimates in
stellar mass.  \citet{mobasher15} estimated the uncertainties
  in their masses to be 0.16~dex, but uncertainties for strong AGNs
  will be larger.
}

\begin{figure}
\epsscale{1.0}
\begin{flushleft}
\plotone{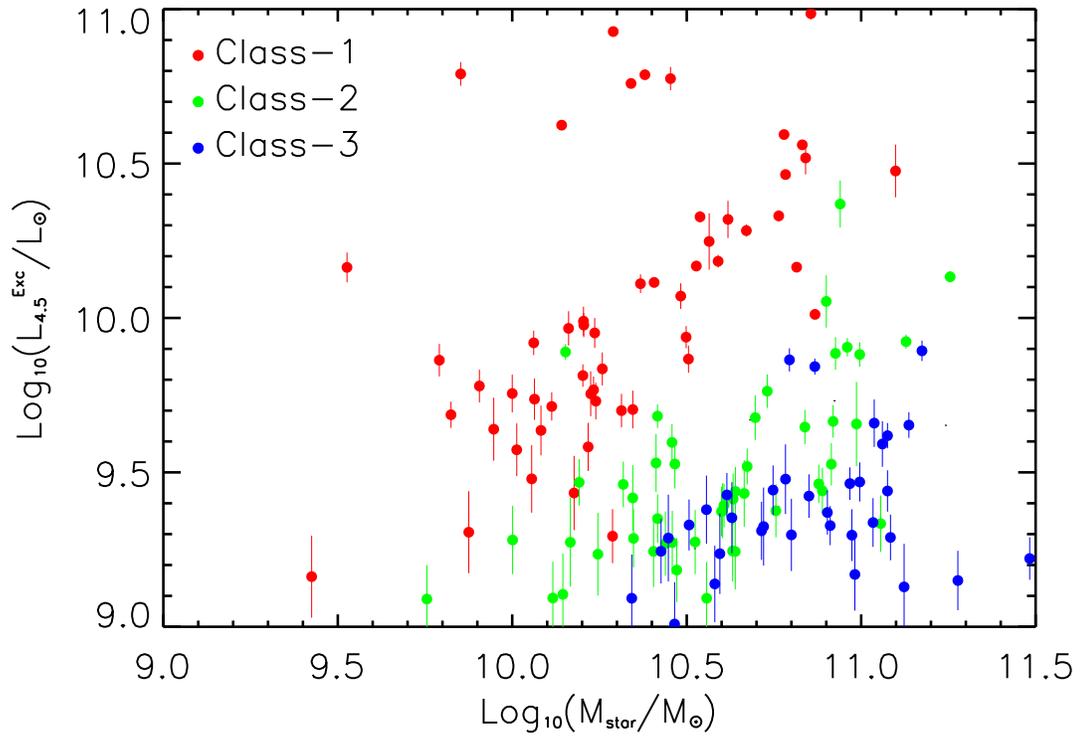}
\end{flushleft}
\caption{AGN luminosity versus stellar mass.  AGN luminosity is here
  measured by $L_{4.5}^{\rm Exc}$, and only galaxies with at least
  3$\sigma$ significant values are shown.  Points are color coded by
  SED Class as indicated in the legend.  }
\label{f:l4p5_mass}
\end{figure}

Figure~\ref{f:l4p5_mass} shows that the AGN luminosity, 
$L_{4.5}^{\rm Exc}$, is not correlated with host galaxy stellar mass.
(For this study, we have
used $L_{4.5}^{\rm Exc}$ instead of $L_X$ as a proxy for AGN 
luminosity.)
The Class~1 subsample (AGNs) has the lowest average stellar mass
{($\log_{10} M_*/\Msun=10.16\pm 0.59$ standard deviation)}, and
objects with Class~3 SEDs (star forming galaxies) have the highest
average stellar mass
{($\log_{10} M_*/\Msun=10.35\pm 0.41$). 
  Values for Class~2 are intermediate $\log_{10} M_*/\Msun=10.29\pm 0.38$.}

\begin{figure}
\begin{flushleft}
\plotone{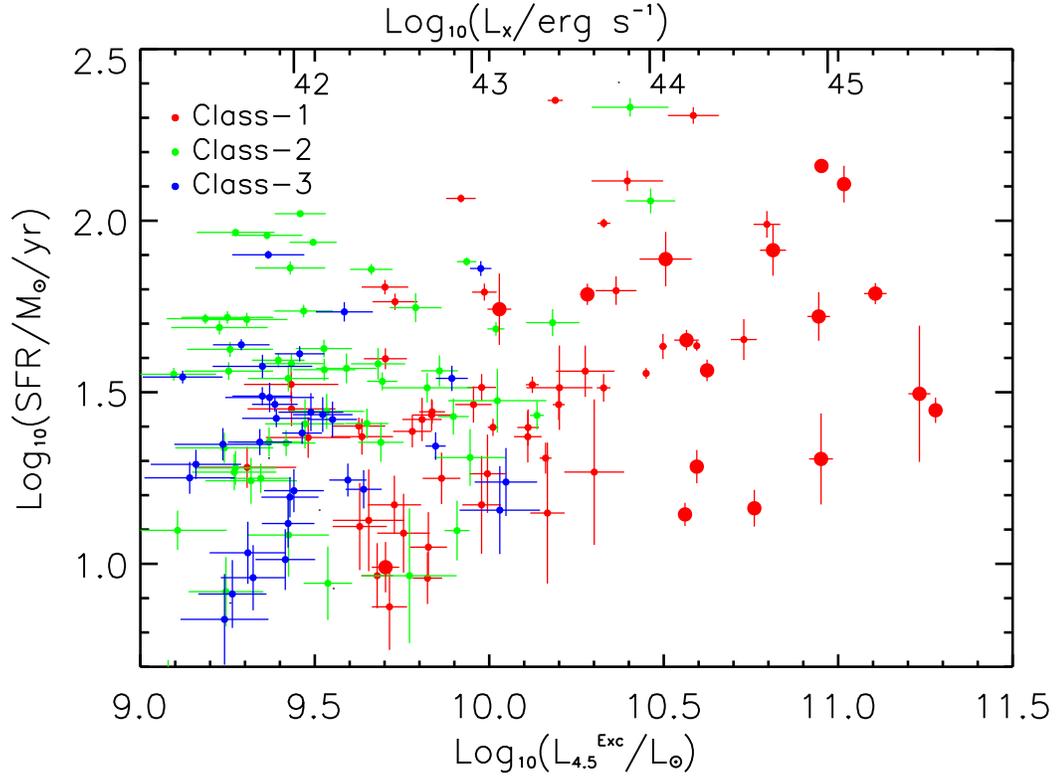}
\end{flushleft}
\caption{AGN luminosity versus SFR.  AGN luminosity is here measured
  by $L_{4.5}^{\rm Exc}$, and only galaxies with at least 3$\sigma$
  significant values are shown. Points are color coded by SED Class
  as indicated in the legend.
  Large points indicate galaxies with low $L_8^{SF}/L_8$ whose
  SFRs were derived from their FIR luminosities. Other SFRs used in
  this plot were derived from $L_8^{\rm SFR}$ (Eq.~A6). The range on
  the abscissa is equivalent to $41<\log_{10}(L_X/\ergs)<44$
  according to the $L_{4.5}^{\rm Exc}$--$L_X$ correlation in
  Fig.~\ref{f:l4.5_lx}, and corresponding values of $L_X$ are shown
  on the upper abscissa. { The Spearman coefficient is 0.07 for all objects in this figure and about 0.2 for objects with Class~1 SEDs.}}
\label{f:l4p5_SFR}
\end{figure}

Figure~\ref{f:l4p5_SFR} shows a positive but weak
correlation between $L_{4.5}^{\rm Exc}$ and SFR for objects with 
Class~1 SEDs {(Spearman rank coefficient $r_s=0.2$)}.   
Previous studies in
this AGN luminosity range
\citep{chen2013,rosario2013,hickox2014,azadi2015,chen2015,xu2015,dai2018}
yielded a wide range of relations between
AGN luminosity and SFR from almost linear to none. At
a given $L_{4.5}^{\rm Exc}$, objects with a Class~2 SED have a higher
SFR than objects with a Class~1 SED.  This is consistent with the
definition of a Class~2 SED as a composite one, consisting of both
strong star formation and AGN features.  There are 32 objects with
significant $L_{4.5}^{\rm Exc}$ and a Class~3 SED, but most are weak
AGNs with $L_{4.5}^{\rm Exc}\la10^{9.5}$\,\Lsun\ and
$L_X\la10^{42}$\,\ergs.  % 3:09pm

\begin{figure}
\begin{flushleft}
\plotone{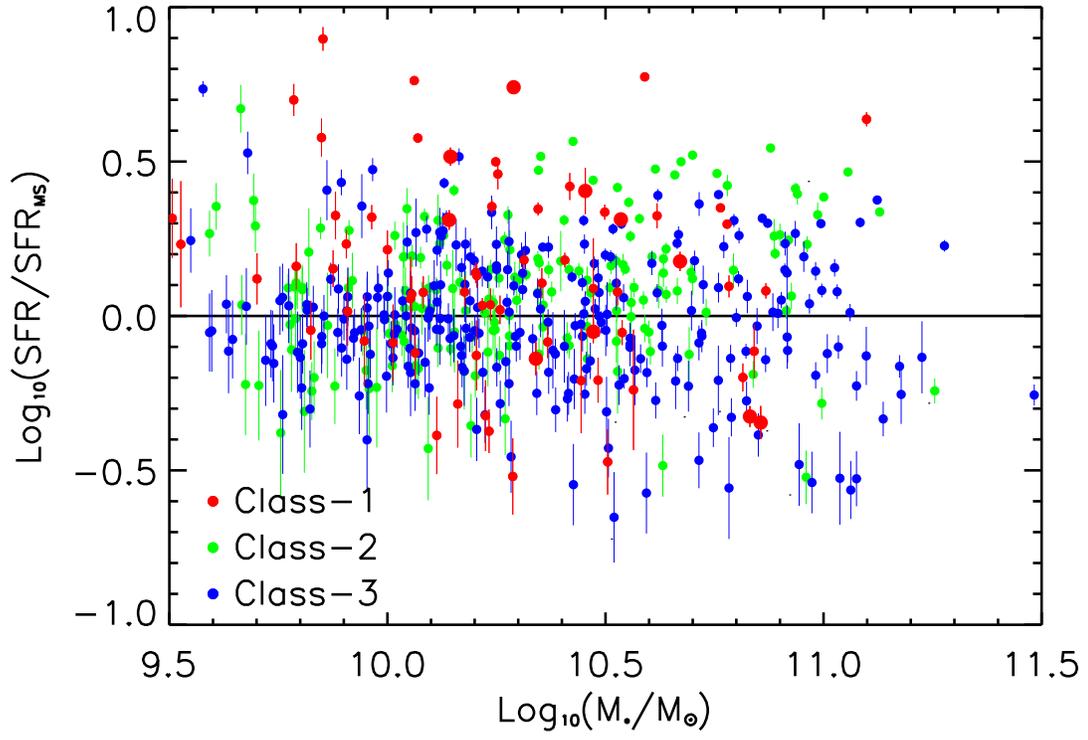}
\end{flushleft}
\caption{SFR relative to the galaxy main sequence for the 16\,\micron-selected
  galaxies.  The main sequence SFR as a function of $M_*$ and redshift was from
  \citet{lee2015}.  Points show galaxies color coded by SED type as
  indicated in the legend, and
  the horizontal line marks the main sequence.  Larger dots indicate
  the objects with strong AGN emission whose SFRs were derived from
  FIR luminosity.  }
\label{f:pah_sfr_ms}
\end{figure}

\begin{figure}
\begin{flushleft}
\plotone{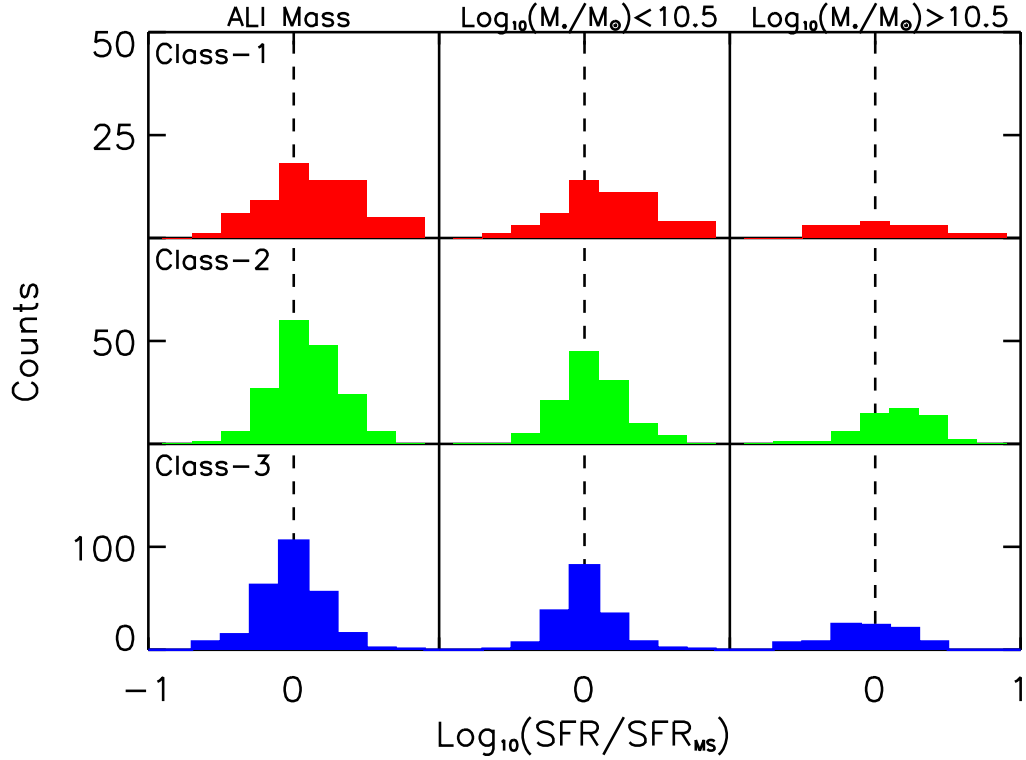}
\end{flushleft}
\caption{Histograms of SFR relative to the galaxy main sequence.  Each row shows one SED Class. Left column shows all objects in the
  respective excess condition (top row) or Class (other rows), and
  middle and right columns show objects 
  with low and high stellar mass as indicated.}
\label{f:pah_sfr_ms_hist}
\clearpage
\end{figure}

Most galaxies at $z\approx1$ have SFR correlated with their stellar
mass; this is known as the main-sequence relation \citep{elbaz2011}.
When star formation is quenched, a galaxy drops off the main-sequence
relation and becomes quiescent.  Several models
\citep[e.g.,][]{ishibashi2012, fabian2012,alatalo2015, man2018,Chen2020}
propose that an AGN is involved in quenching the star
formation. Figure~\ref{f:pah_sfr_ms} shows SFR relative to the
main-sequence SFR at each galaxy's redshift.  Most galaxies scatter around the
main sequence regardless of their SED Class.  This is further
demonstrated in Figure~\ref{f:pah_sfr_ms_hist}, which shows that most
galaxies, including those with AGNs, have SFR on the main
sequence. Indeed a majority of galaxies with $M_*>10^{10.5}$\,\Msun\
and Class~1 or Class~2 SEDs have SFR above the main sequence despite
all AGNs having a symmetric distribution around the main
sequence. This is generally consistent with previous observations
that X-ray-selected AGNs up to $z\sim4$ are on the main sequence
\citep[e.g.,][]{mullaney2015, aird2019,bernhard2019, suh2019}.  Our
sample does not include a substantial AGN population below the main
sequence as found by some \citep{shimizu2015, bain2020, stemo2020}
for local and for distant but $z<4$ AGNs.  Presumably this is a result of our
rest $\sim$8\,\micron\ sample selection. The SED Class of each AGN host in our
sample is mainly controlled by AGN luminosity. Galaxies with higher AGN
luminosities tend to fall into Class~1 because the AGN overwhelms
star-formation signatures.  The overall tendency of galaxies in all
Classes to follow the main-sequence relation suggests that AGN
luminosity and SFR change in tandem.

\begin{figure}
\begin{flushleft}
\plotone{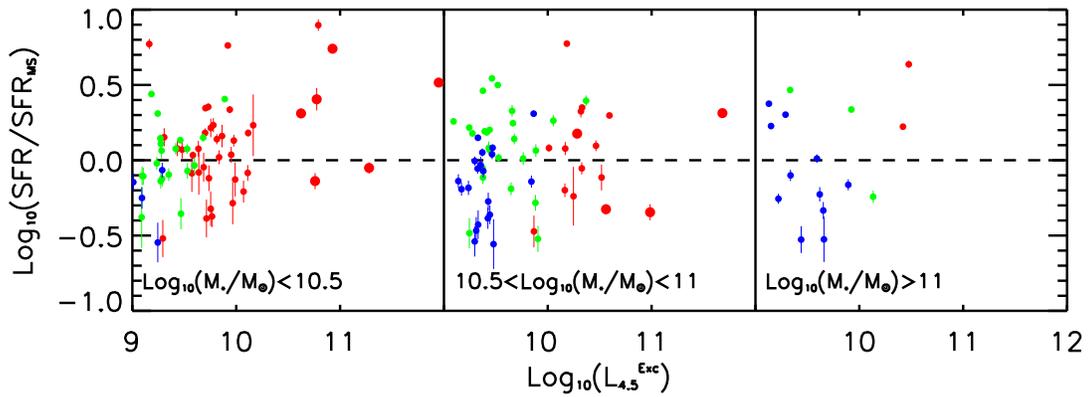}
\end{flushleft}
\caption{SFR relative to the galaxy main sequence versus AGN
  luminosity.  AGN luminosity is here measured by $L_{4.5}^{\rm
    Exc}$, and only galaxies with at least 3$\sigma$ significant
  values are shown. Points are color coded by SED type: red for
  Class~1, green for Class~2, and blue for Class~3. Panels show three
  bins of stellar mass as labeled. }
\label{f:pah_sfr_norm_l4p5}
\end{figure}

Galaxies in our sample show no reduction in SFR when an AGN is
present, as one might expect in an AGN-quenching scenario
\citep{fabian2012,ishibashi2012, alatalo2015, man2018}.
Figure~\ref{f:pah_sfr_ms_hist} shows that galaxies with an AGN
component show a similar
SFR/SFR$_{\rm MS}$ distribution to that of star-forming galaxies.
Indeed an AGN component could exist in most galaxies in our
sample, though it would be veiled in the ones with a higher SFR. On
the other hand, galaxies with Class~1 SEDs, along with high-mass
galaxies having Class~2 SEDs, tend to lie above the main
sequence.  This implies if anything enhanced star formation in
AGN-dominated galaxies.  To further test this,
Figure~\ref{f:pah_sfr_norm_l4p5} tracks how SFRs are affected by AGNs
in each mass bin.  Class~1 SEDs become rarer as stellar masses
increases, consistent with increasing dominance of the star-formation
component in massive galaxies.  In addition, very few AGNs with
$\log_{10}(L_{4.5}^{\rm Exc}/\Lsun)>10$ have SFR below the main
sequence value.  This is particularly noticeable for AGNs in the
$10.5<\log_{10}(M_*/\Msun)<11$ bin.  One interpretation of
Figure~\ref{f:pah_sfr_norm_l4p5} is that AGN luminosity decreases
when its host galaxy evolves across the main sequence line.  In the
highest-mass $\log_{10}(M_*/\Msun)>11$ bin, all AGNs have lower
$L_{4.5}^{\rm Exc}$, and only three of them are above the main
sequence. The implication on the distribution of AGNs along the main
sequence is two-fold: (1) AGN accretion increases together with the
star formation activity; (2) whatever quenches the star formation may
also quench its central AGN accretion. 
Given the very different timescales involved,
it is difficult to draw conclusions about AGN feedback 
as a quenching mechanism from the comparison of AGN emission and IR
SFRs in our sample, but
the parallel decline in AGN accretion rate and SFR
seen here is consistent with
so-called halo-quenching models, in which halo gas cooling becomes
less effective as halo mass increases
\citep[e.g.,][]{Correa2018}.  
The decline in global cool-gas inflow  thus deprives
both the galaxy and the black hole of further fuel to grow.

\section{SUMMARY}

A sample of 705  16\,\micron-selected, $0.8<z<1.3$ galaxies
mostly have $11<\log_{10}(\lir/\Lsun)<12$, qualifying them as LIRGs. 
{In most cases,} their
rest-frame $1.6\,\micron<\lambda<12$\,\micron\ SEDs are well fit
by local galaxy SED templates.  Most galaxies in the sample follow
the galaxy main sequence relation between stellar mass and SFR. Their
8\,\micron\ luminosity function shows a strong evolution from
$z\sim0$ corresponding to the strong SFRD evolution over that
redshift range.

Based on fitting the MIR SEDs, {84\% of the galaxies in our
16\,\micron\ sample are star forming: best fit by either star-forming
or composite templates.}  While all of
these are likely to show PAH emission and be forming stars,
{about 17\% of the Class~2+3 galaxies}
show an AGN component revealed by either
{4.5\,\micron\
luminosity exceeding that from stellar processes (stellar
photospheres plus dust heated by
star formation) or X-ray luminosity exceeding 10$^{42}$\,\ergs.}
About 15\% of the sample are best
fit by an AGN template. 
No object in our sample can be fit by a quiescent galaxy template, not
surprising given the 16\,\micron\ selection.  Nine objects (1.3\%)
show neither PAH emission nor strong rest-4.5\,\micron\ continuum.
All of these objects 
have $M_*>10^{10}$\,\Msun\ although the local templates that fit
their SEDs have $M_*<10^9$\,\Msun. Further investigation is needed
for these special objects.

Our fitting of galaxy SEDs for the sample permits accurate separation
of MIR luminosities contributed by star formation and an AGN
component.  This is because the $L_{4.5}/L_{1.6}$ and $L_{8}/L_{1.6}$
color--color diagram shows a close correlation for star-forming
galaxies, but galaxies with an AGNs show excess emission at
4.5\,\micron\ ($L_{4.5}^{\rm Exc}$).  This excess correlates well
with X-ray luminosities for galaxies having \Ch\ X-ray detections,
and $L_{4.5}^{\rm Exc}$ should therefore be a useful AGN luminosity
indicator, as it is for local galaxies. As to SFR indicators, SFRs estimated from 8\,\micron\
luminosity ($L_8$), 24\,\micron\ flux density, FIR luminosity, and
UV-corrected flux yield consistent results for this sample of
LIRGs. SFRs derived from $L_8$ and 24\,\micron\ flux density have the
lowest scatter.  Therefore we suggest that $L_8$ can trace the SFR as
accurately as the 24\,\micron\ flux density for this LIRG sample and more
accurately than SFR derived from UV flux.

Our sample galaxies show a correlation between AGN luminosity (as
measured by $L_{4.5}^{\rm Exc}$) and SFR, indicating a coevolution
between black hole accretion rate and star formation. Also, galaxies
with high AGN luminosities and an SED showing a dominant AGN tend to
lie above the star formation main sequence.  These galaxies are seen
predominantly at low stellar mass and are absent at
$M_*>10^{11}$\,\Msun.  AGN accretion therefore appears to shut down
along with star formation when enough stellar mass has been
accumulated.  Although galaxies with and without AGN follow the main
sequence, we cannot rule out the possibility of AGN-related
quenching.  Nevertheless, the positive correlation between AGN and
SFR and the lack of AGNs at the massive end tends to favor the
halo-mass quenching mechanism, which stops not only galaxy-wide star
formation but also gas feeding the accretion of the central massive
black hole.

\acknowledgments {This work is based in part on observations made
  with the \sst, which is operated by the Jet Propulsion Laboratory,
  California Institute of Technology under a contract with
  NASA. Support for this work was provided by the Chinese National
  Nature Science foundation grant number 10878003. This work was
  supported in part by the National Key R\&D Program of China via
  grant number 2017YFA0402703 and by NSFC grants 11433003, 11933003,
  11373034. Dr.\ C.~Cheng is supported by NSFC grants 11803044 and
  11673028. Additional support came from the Chinese Academy of
  Sciences (CAS) through a grant to the South America Center for
  Astronomy (CAS-SACA) in Santiago, Chile. }

%Facilities:
\facilities{Spitzer(IRAC), Spitzer(MIPS), Spitzer(IRS), Keck, Subaru}
\clearpage

\begin{figure} 
\epsscale{0.8}
\plotone{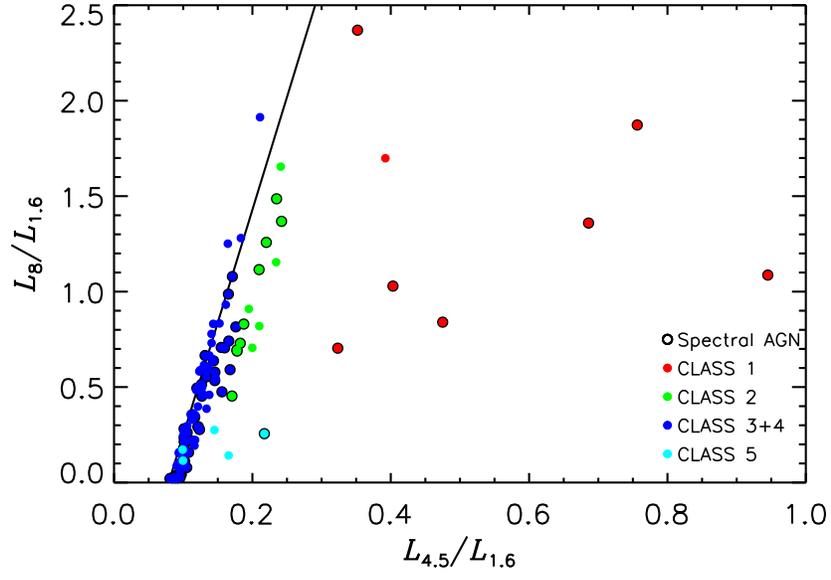}
\caption{MIR--NIR color--color diagram for the template set.  The
  ratio $L_8/L_{1.6}$ roughly traces the specific star formation
  rate, while $L_{4.5}/L_{1.6}$ basically traces the AGN luminosity
  per stellar mass.  Template types are shown by colors as indicated
  in the legend.  The line shows the best-fit relation for
  star-forming templates (Eq.~\ref{eq:nirmir}).}
\label{f:tmp_cc}
\end{figure}

\appendix
\renewcommand{\theequation}{A\arabic{equation}}
Measured broadband (rest-frame) luminosities $L_8$ and $L_{4.5}$ can
be considered as the sum of star-forming and AGN components:
\begin{equation}
L_8=L_8^{\rm SFR}+L_8^{\rm Exc}\quad,
\end{equation}
and
\begin{equation}
L_{4.5}=L_{4.5}^{\rm SFR}+L_{4.5}^{\rm Exc}\quad.
\end{equation}
Here $L_{4.5}^{\rm SFR}$ and $L_8^{\rm SFR}$ consist of both
photospheric emission from stars and dust emission
from star-formation regions, and $L_{4.5}^{\rm Exc}$ is taken to be
AGN luminosity. For both $L_{4.5}^{\rm SFR}$ and $L_8^{\rm SFR}$, we
adopt the relation found in Figure~\ref{f:tmp_cc}, and  we assume
that $L_{1.6}$ is only from stellar photospheres:
\begin{equation}
L_{4.5}^{\rm SFR}=(L_8^{\rm SFR}+0.93L_{1.6})/11.8\quad.
\end{equation}
The AGN intrinsic SED model from \citet{mullaney2011} gives a power law
$\lambda F_{\lambda}\propto\lambda^{\alpha}$ 
with $-0.3<\alpha<0.8$ for the AGN component. This gives
\begin{equation}
 L_8^{\rm Exc}/L_{4.5}^{\rm Exc} =(8/4.5)^{\alpha}\quad.
\end{equation}
By solving the above three equations, we reach
\begin{equation}
L_{4.5}^{\rm Exc}=\epsilon(11.8L_{4.5}-L_8+0.93L_{1.6})\quad,
\end{equation}
where $\epsilon=[11.8-(8/4.5)^{\alpha}]^{-1}$, a weak function of
$\alpha$.    For the  expected $\alpha\approx0$,
$\epsilon\approx0.093$. ($\epsilon=0.091$--0.098 for the range of
$\alpha$ above.) The star forming component in $L_8$ is then 
\begin{equation}
L_8^{\rm SFR}=L_8 - L_{4.5}^{\rm Exc}(8/4.5)^\alpha\quad.
\end{equation}
Qualitatively, $L_8$ measures SFR with a modest AGN correction while
$L_{4.5}$ measures the AGN with a modest star-formation
correction. For this LIRG sample, photospheric emission at  rest
8\,\micron\ is only 2--3\% of $L_8^{\rm SFR}$. We therefore need not subtract
photospheric emission  from $L_8$ when calculating
SFR{, and the relatively small coefficient on
$L_{1.6}$ shows that any AGN contribution at 1.6\,\micron\ will make
little difference.} Error bars{, including uncertainties in the parameters,} for $L_8^{\rm SFR}$ and $L_{4.5}^{\rm Exc}$ were
propagated according to the above equations.



\nonfrenchspacing


\clearpage

% 12:50- 1:50pm for captions

\clearpage

\clearpage

\clearpage

\clearpage

\clearpage

\clearpage

\clearpage

\clearpage

\clearpage
\begin{figure}
\epsscale{1.05}
\plotone{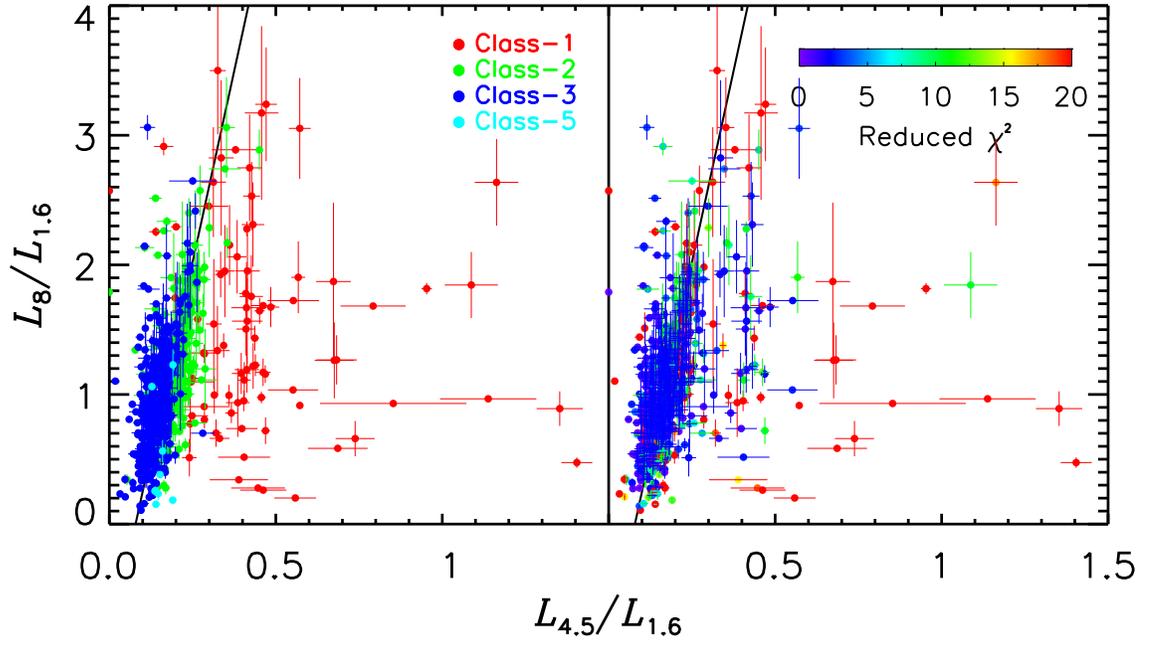}
\caption{Color--color diagram for the 16\,\micron-selected sample
  galaxies.  Points in the left panel are color coded by SED type as
  indicated in the legend. Colors in the right panel indicate reduced
  $\chi^2$ as indicated in the color bar.  The line shows the
  color--color relation for the Class~3 and~4 templates from
  Fig.~\ref{f:tmp_cc}.} 
\label{f:sample_l8l4p5}
\end{figure}
\clearpage

\clearpage

\clearpage

\clearpage

\clearpage

\clearpage

\clearpage

\clearpage

\clearpage

\clearpage

\clearpage

\clearpage

\clearpage

\clearpage

\clearpage

\clearpage

\begin{table}[hbt]
{%\scriptsize
\begin{center}
\vspace{0.1cm}
\caption{\sc\protect\centering  The 16\,\micron-Selected Sample }
\vspace{0.3cm}
\begin{tabular}{lccccc}
\hline\hline \noalign{\smallskip}
Field &  Area       & $f_{\rm limit}(16\,\micron)$ & \# of &  \# of
$z_{\rm spec}$& redshift references\tablenotemark{a} \\
      &  arcmin$^2$ &   \mmjy     &   objects     
  & & $z_{\rm spec}$, $z_{\rm phot}$\\
\hline
\noalign{\smallskip}
EGS     & 432 & 130  & 263  &   190&\citet{Newman2013}, \citet{huang2013} \\
GOODS-N & 158 & 30   & 334  &   264&\citet{Wirth2004}, \citet{dahlen2013}\\
GOODS-S & 104 & 30   & 108  &   102&\citet{Balestra2010}, \citet{dahlen2013}\\
\hline
total   & 694 &      & 605  &   556&\\
\noalign{\hrule}
\noalign{\smallskip}
\label{tab:sample}
\end{tabular}
\tablenotetext{a}{Major surveys are indicated, but additional
  redshifts were collected from other publications.}
\end{center}
}
\end{table}

%\clearpage

\begin{table}[hbt]
{%\scriptsize
\begin{center}
\vspace{0.1cm}
\caption{\sc\protect\centering 8\,\micron\  Luminosity Function
  Function at $z\approx1$} 
\vspace{0.3cm}
\begin{tabular}{ccc}
\hline\hline \noalign{\smallskip}

$\log L_8/\Lsun$ & 
$\log( \phi$)\tablenotemark{a} &$\delta\log(\phi$)\tablenotemark{a} \\
\hline \noalign{\smallskip} 

\09.8 & $-$2.95 & 0.43 \cr
10.0 & $-$2.29 & 0.13 \cr
10.2 & $-$2.47 & 0.07 \cr
10.4 & $-$2.43 & 0.05 \cr
10.6 & $-$2.33 & 0.04 \cr
10.8 & $-$2.43 & 0.04 \cr
11.0 & $-$2.76 & 0.05 \cr
11.2 & $-$3.21 & 0.09 \cr
11.4 & $-$4.01 & 0.17 \cr
11.6 & $-$4.66 & 0.24 \cr

\noalign{\hrule}
\noalign{\smallskip}
\label{tab:lf8}
\end{tabular}
\tablenotetext{a}{Units of $\phi$ are galaxies per comoving Mpc$^{-3}$ for
  $H_0=70$\,\kms\,Mpc$^{-1}$.} 
\end{center}
}
\end{table}

%\clearpage

\begin{table*}[hbt]
{%\scriptsize
\begin{center}
\vspace{0.1cm}
\caption{\sc\protect\centering Template Classification Criteria}
\vspace{0.3cm}
\begin{tabular}{llcc}
\hline\hline \noalign{\smallskip}
Template & type & 6.2\,\micron\  PAH EW & $L_8/L_{1.6}$\\
\hline
Class~1 & AGN & $   $ & $5\times\,L_{4.5}/L_{1.6}-EW>1.145$\\
Class~2 & Composite & ${\rm EW}>0.1$ & $1.145>5\times\,L_{4.5}/L_{1.6}-EW>0.445$\\
Class~3
        & Star Forming & ${\rm EW}>0.1$   &  $5\times\,L_{4.5}/L_{1.6}-EW<0.445$   \\
Class~4 & Quiescent & $\rm EW=0$ & $L_{4.5}/L_{1.6}\sim0.1$\\
Class~5 & Blue Compact & ${\rm EW}<0.1$ & $L_{4.5}/L_{1.6}<0.25$  \\
\noalign{\hrule}
\noalign{\smallskip}
\label{tab:criteria}
\end{tabular}
\end{center}
}
\end{table*}
%\clearpage

\begin{table*}[hbt]
{%\scriptsize
\begin{center}
\vspace{0.1cm}
\caption{\sc\protect\centering SED Classes for the 16\,\micron\
  Selected Sample } 
\vspace{0.3cm}
\begin{tabular}{lcccc}
\hline\hline \noalign{\smallskip}
Field &  Class~1      & Class~2 & Class~3 &  Class~5\\
      &  AGN &   Composite     & Star-forming      &  Blue Compact \\
\hline
\noalign{\smallskip}
EGS     & 43 & 78  & 142  &   0\\
GOODS N & 49 & 115  & 161  &   9\\
GOODS S & 15 & 30   & 63  &   0\\
\noalign{\hrule}
All     & 107& 223  & 366 & 9 \\
Fractions & 15.2\% & 31.6\% & 51.9\% & 1.3\% \\
\noalign{\hrule}
\noalign{\smallskip}
\label{tab:SED_fitting}
\end{tabular}
\end{center}
}
\end{table*}

\begin{table}[hbt]
{%\scriptsize
\begin{center}
\vspace{0.1cm}
\caption{\sc\protect\centering Number of Each
  SED Class Detected in X-rays } 
\vspace{0.3cm}
\begin{tabular}{lrrrr}
\hline\hline \noalign{\smallskip}
Field & \Ch        &Class~1 & Class~2 &  Class~3  \\
      & exp.\ time & AGN &   Composite     &   Star-forming  \\
\hline
\noalign{\smallskip}
\multicolumn{5}{l}{all \Ch\ X-ray detections}\\
EGS     & 0.8 Ms&  7 & 10 &  7\\
GOODS-N &   2 Ms& 22 & 17 & 16\\
GOODS-S &   7 Ms& 12 & 10 & 24\\
\hline
\multicolumn{5}{l}{with $L_x>10^{42}$\,\ergs}\\
EGS     &       &  7 & 10  &  7\\
GOODS-N &       & 20 & 9  &  10\\
GOODS-S &       &  4 & 2  &  2\\
\hline
%Total   & 13.5& 45.8 &39.3\\
\noalign{\hrule}
\noalign{\smallskip}
\label{tab:X-sample}
\end{tabular}
\tablecomments{\protect\raggedright
    \citet{xue2016} and \citet{luo2017} reported X-ray
    luminosities for GOODS-N and GOODS-S respectively for the energy
    band 0.7--7~keV.  To be consistent with EGS luminosities, we have
    converted to luminosities in the 2--10\,keV band by multiplying
    by 0.721.}
%We adopted $L_{\rm 2-10\,keV}$ as $L_X$ and converted $L_{\rm 0.5-7\,keV}$ in 
%\citet{xue2016,luo2017} to $L_X$ with $L_X=0.721L_{\rm 0.5-7\,keV}$. 

%\tablenotetext{a}{with $L_x>10^{42}$\,\ergs}
\end{center}
}
\end{table}

\end{document}